\documentclass[a4paper,fleqn,usenatbib]{mnras}

\usepackage{newtxtext,newtxmath}
\usepackage[T1]{fontenc}
\usepackage{ae,aecompl}

\usepackage{graphicx}	% Including figure files
\usepackage{amsmath}	% Advanced maths commands
\usepackage{amssymb}	% Extra maths symbols

\title[Resolved cm-wavelength observations of MWC\,758]{Cm-wavelength
  observations of MWC\,758: resolved dust trapping in a vortex}

\author[S. Casassus et al.]{
  Simon Casassus,$^{1,2}$\thanks{E-mail: simon@das.uchile.cl}
  Sebastian Marino, $^{3}$  
  Wladimir Lyra, $^{4,5}$
  Cl\'ement Baruteau, $^{6}$    \newauthor
  Mat\'{\i}as Vidal, $^{1}$  
  Alwyn Wootten, $^{7}$  
  Sebasti\'{a}n P\'{e}rez, $^{1}$ 
  Felipe Alarcon, $^{1}$  \newauthor  
  Marcelo Barraza, $^{1}$  
  Miguel C\'arcamo, $^{8}$  
  Ruobing Dong, $^9$
  Anibal Sierra,$^{10}$
  Zhaohuan Zhu, $^{11}$   \newauthor
  Luca Ricci, $^{4}$ 
  Valentin Christiaens, $^{1}$
  Lucas Cieza, $^{12}$
\\
$^{1}$Departamento de Astronom\'{\i}a, Universidad de Chile, Casilla 36-D, Santiago, Chile\\
$^{2}$Millennium Nucleus ``Protoplanetary Disks'', Chile\\
$^{3}$Institute of Astronomy, University of Cambridge, Madingley Road, Cambridge CB3 0HA, UK\\
$^{4}$Department of Physics and Astronomy, California State University Northrige, 18111 Nordhoff St, Northridge CA 91130, USA\\
$^{5}$Jet Propulsion Laboratory, California Institute of Technology, 4800 Oak Grove Drive, Pasadena, CA, 91109, USA\\
$^{6}$CNRS / Institut de Recherche en Astrophysique et Plan\'etologie, 14 avenue Edouard Belin, 31400 Toulouse, France\\
$^{7}$National Radio Astronomy Observatory, Charlottesville, VA 22903, USA\\
$^{8}$Departamento de Ingenier\'{\i}a Inform\'atica, Universidad de Santiago de Chile\\
$^{9}$Department of Physics \& Astronomy, University of Victoria, Victoria, BC, V8P 1A1, Canada\\
$^{10}$Instituto de Radioastronom\'{\i}a y Astrof\'{\i}sica, UNAM, Apartado
Postal 3-72, 58089 Morelia Michoac\'an, M\'exico\\
$^{11}$Physics \& Astronomy University of Nevada - Las Vegas\\
$^{12}$Facultad de Ingenier\'{\i}a y Ciencias, N\'ucleo de Astronom\'{\i}a, Universidad Diego Portales, Av. Ejercito 441. Santiago, Chile
}

\date{Accepted XXX. Received YYY; in original form ZZZ}

\pubyear{2017}

\begin{document}
\label{firstpage}
\pagerange{\pageref{firstpage}--\pageref{lastpage}}
\maketitle

% Abstract of the paper  250 words max
\begin{abstract}

The large crescents imaged by ALMA in transition disks suggest that
azimuthal dust trapping concentrates the larger grains, but
centimetre-wavelengths continuum observations are required to map the
distribution of the largest observable grains. A previous detection at
$\sim$1\,cm of an unresolved clump along the outer ring of MWC\,758
(Clump\,1), and buried inside more extended sub-mm continuum,
motivates followup VLA observations. Deep multiconfiguration
integrations reveal the morphology of Clump 1 and additional cm-wave
components which we characterize via comparison with a deconvolution
of recent 342~GHz data ($\sim$1\,mm). Clump\,1, which concentrates
$\sim1/3$ of the whole disk flux density at $\sim$1\,cm, is resolved
as a narrow arc with a deprojected aspect ratio $\chi > 5.6$, and with
half the azimuthal width than at 342 GHz. The spectral trends in the
morphology of Clump\,1 are quantitatively consistent with the Lyra-Lin
prescriptions for dust trapping in an anticyclonic vortex, provided
with porous grains ($f \sim 0.2 \pm 0.2$) in a very elongated ($\chi
\sim 14 \pm3$) and cold ($T\sim23\pm2\,$K) vortex. The same
prescriptions constrain the turbulence parameter $\alpha$ and the gas
surface density $\Sigma_g$ through $\log_{10}\left( \alpha \times
\Sigma_g / \mathrm{g\,cm}^{-2} \right) \sim -2.3\pm0.4$, thus
requiring values for $\Sigma_g$ larger than a factor of a few compared
to that reported in the literature from the CO isotopologues, if
$\alpha \lesssim 10^{-3}$. Such physical conditions imply an
appreciably optically thick continuum even at cm-wavelengths
($\tau_{33\,\mathrm{GHz}}\sim 0.2$).  A secondary and shallower peak
at 342\,GHz is about twice fainter relative to Clump\,1 at
33\,GHz. Clump\,2 appears to be less efficient at trapping large
grains.

\end{abstract}

%. In summary, the new VLA data show that Clump\,1 is likely due to the
%pile-up of fluffy and large grains, and is thus an ideal source to
%observe grain growth or grain size segregation.

\begin{keywords}
%stars: variables: T Tauri, Herbig Ae/Be --  protoplanetary discs --- stars: formation -- accretion, accretion discs -- planet–disc interactions
protoplanetary discs ---  accretion, accretion discs  --- planet-disc interactions
\end{keywords}

%%%%%%%%%%%%%%%%%%%%%%%%%%%%%%%%%%%%%%%%%%%%%%%%%%

%%%%%%%%%%%%%%%%% BODY OF PAPER %%%%%%%%%%%%%%%%%%

\section{Introduction}

A pathway to the formation of planetesimals, and eventually giant
planets, may occur in compact concentrations of dust grains trapped in
pressure maxima \citep[][]{Weidenschilling1977MNRAS.180...57W,
  Cuzzi2008ApJ...687.1432C}. The pile-up of the larger grains
\citep[][]{Barge_Sommeria_1995A&A...295L...1B,
  Birnstiel2013A&A...550L...8B, LyraLin2013ApJ...775...17L,
  Zhu_Stone_2014ApJ...795...53Z, Mittal2015ApJ...798L..25M,
  BZ2016MNRAS.458.3927B}, which would otherwise rapidly migrate
inwards due to aerodynamic drag
\citep[][]{Weidenschilling1977MNRAS.180...57W}, could lead to the
genesis of planet embryos \citep[][]{Lyra2009A&A...497..869L,
  Sandor2011ApJ...728L...9S}.

%The aerodynamic coupling of gas and dust is bound to bear a strong
%impact in circumstellar disk evolution.

The observational identification of so-called dust traps in the form
of large-scale crescents of mm-wavelength-emitting dust grains
\citep[][mm-grains for short]{Casassus2013Natur,
  vanderMarel2013Sci...340.1199V, Perez_L_2014ApJ...783L..13P},
suggests that azimuthal dust trapping has major structural
consequences in protoplanetary disks. In the dust trapping paradigm to
explain the large crescents, the origin of the pressure maximum itself
is unknown, and could for example be due to anticyclonic vortices,
which could be induced by the formation of a planetary gap
\citep[][]{Zhu_Stone_2014ApJ...795...53Z, Koller2003ApJ...596L..91K,
  deVal-Borro2007A&A...471.1043D}, or by discontinuities in the disk
viscosity \citep[e.g. at the edge of a dead
  zone,][]{Varniere2006A&A...446L..13V, Regaly2012MNRAS.419.1701R}.
Simulations with consistent disk self-gravity
\citep[][]{ZB2016MNRAS.458.3918Z}, suggest that the contrast ratio
between the maximum and minimum along the outer ring gas surface
density can reach $\sim$3, at most, for either a planetary gap or a
dead zone.  Recent advances in hydrodynamic simulations of
circumbinary disks have, however, produced very lopsided gas rings
with contrast ratios of $\gtrsim$10, and negligible azimuthal dust
trapping for mm-grains \citep[][]{Ragusa2017MNRAS.464.1449R}.

Thus the more pronounced contrast ratios seen in the continuum, of
$\sim$30 in HD\,142527 \citep[][]{Casassus2013Natur,
  Casassus2015ApJ...812..126C, Muto2015PASJ...67..122M,
  Boehler2017ApJ...840...60B} and $\sim$\,100 in IRS\,48
\citep[][]{vanderMarel2013Sci...340.1199V,vandermarel_2015ApJ...810L...7V},
have been interpreted as likely due to dust trapping in a vortex
\citep[e.g.][]{LyraLin2013ApJ...775...17L, BZ2016MNRAS.458.3927B,
  Sierra2017ApJ...850..115S}. But, while in HD\,142527 the evidence
from the multi-frequency dust continuum alone would suggest that
trapping likely occurs for larger cm-sized grains, and not for
mm-sized grains \citep[][]{Casassus2015ApJ...812..126C}, the required
lopsided gas ring is not observed in CO isotopologues
\citep[][]{Muto2015PASJ...67..122M, Boehler2017ApJ...840...60B}.
Crescents with extreme contrasts such as in HD\,142527 and IRS\,48 are
however rarely observed, these two sources being examples of the very
brightest protoplanetary disks \citep[and HD\,142527 can be seen as a
  circumbinary disk, e.g.][]{Biller2012,
  Christiaens2018A&A...617A..37C}.  Large crescents with smaller
contrast ratios are nonetheless common in the sub-mm continuum from
the outer rings of protoplanetary disks with large central cavities
(i.e. so-called transition disks), as in LkH$\alpha$330
\citep[][]{Isella2013ApJ...775...30I}, SR\,21, HD135344B
\citep[][]{Perez_L_2014ApJ...783L..13P,
  vandermarel_2015A&A...579A.106V, vdM2016ApJ...832..178V}, DoAr\,44
\citep[][]{vdMarel_2016A&A...585A..58V}, and HD\,34282
\citep[][]{vdP2017A&A...607A..55V}.

%Taking the above scenario for azimuthal dust trapping one step further
%poses a question on the potential of such dust traps to form gaseous
%giants. What would be the impact on the disk of massive bodies forming
%inside a dust trap at large stellocentric distances? Are there signs
%of such bodies?

Smaller clumps of cm-wavelength continuum emission are another type of
azimuthal structure observed in transition disks. Clumpy rings have
been seen in, for example, HL\,Tau, HD\,169142, and LkCa 15
\citep[][]{ CarrascoGonzalez2016ApJ...821L..16C,
  Macias2017ApJ...838...97M, Isella2014ApJ...788..129I}, although
further observations are required to ascertain the significance of the
clumpy structure. However, an example stands out as an intriguing
radio continuum clump atop more extended emission: the unresolved
34\,GHz signal detected by \citet{Marino2015ApJ...813...76M} in
MWC\,758 using the NSF's Karl G. Jansky Very Large Array (VLA), in B
array. This clump encloses a few Earth masses in dust. The clump
recently reported in HD\,34282 by \citet[][]{vdP2017A&A...607A..55V}
bears similarities with MWC\,758, considering that it is more extended
as seen in the sub-mm continuum.

%Another similar clump has recently been reported in HD\,34282
%\citep[although more extended as seen in the sub-mm
%  continuum,][]{vdP2017A&A...607A..55V}.

MWC\,758, at a distance of $160.2\pm1.7$\,pc
\citep[][]{Gaia2016A&A...595A...1G}, is also a Herbig disk viewed
close to face-on, as is HD\,142527, although its cavity is not as deep
in scattered light \citep[][]{Grady2013ApJ...762...48G,
  Benisty2015A&A...578L...6B}. The preliminary VLA observations
\citep[][VLA/13B-273]{Marino2015ApJ...813...76M}, with 1\,h on-source,
revealed an unresolved clump to the North. The VLA emission is more
concentrated compared to the higher frequency band\,7 data obtained
with the Atacama Large Millimeter/submillimeter Array (ALMA).  Even
after convolution to the coarser ALMA beam, the area inside the 0.85
intensity maximum contour in the VLA map is 0.09\arcsec$^2$, while it
is 0.23\arcsec$^2$ in the ALMA map. It would thus seem that either
azimuthal dust trapping is at work in MWC\,758, so  the larger
cm-wavelength emitting grains (cm-grains for short) are trapped more
efficiently, or that the VLA continuum pierces through an optically
thick 850\,$\mu$m continuum. Indeed,
\citet{Boehler2018ApJ...853..162B} reported higher angular resolution
ALMA observations that confirm this two-clump structure, which they
model with significant increases in the dust-to-gas mass ratio,
consistent with the dust trap origin. Band\,7 continuum observations
with finer yet angular resolutions have recently been reported by
\citet{Dong2018ApJ...860..124D}

%CO LINES DEV DEV 
%In turn, the available CO(3-2) data (Fig.\,\ref{fig:velocity}) also
%point at a companion. The velocity field is roughly Keplerian, but
%after accounting for the finite beam, there is additional fine
%structure in the centroid map near the VLA clump. Such kinematical
%features may be related to circumplanetary disks
%\citep{Perez2015ApJ...811L...5P}. The integrated intensity is centered
%on the star, so that this is the main source of CO excitation and any
%companion is therefore not massive enough to heat their surroundings
%as a star would The CO disk is abruptly truncated at $\sim$1.5\arcsec,
%so approximately 3$\times$ a binary separation of 0.5\arcsec, and
%suggestive of dynamical
%interactions\citep{Artymowicz_Lubow_1994ApJ...421..651A}.

Here we followup the preliminary detection of a compact dust
concentration in MWC\,758 with deep integrations in VLA A, B and C
configurations (Sec.\,\ref{sec:obs}). A comparison with re-processed archival
ALMA observations suggests that the VLA clump lies embedded within a
sub-mm arc-like structure (Sec.\,\ref{sec:ALMA_VLA}). We quantify
spectral trends in terms of the arc lengths and aspect ratios, and
show that they are quantitatively consistent with the dust trapping
scenario (Sec.\,\ref{sec:discussion}).  We conclude on the main
features observed in MWC\,758, and on their connection with the dust
trapping scenario (Sec.\,\ref{sec:conclusion}).

\section{New VLA observations} \label{sec:obs}

%LOG
%C-array
%strelka12:53:21/data/simon/VLA_MWC758/C-array/16A-314_2016_02_14_T06_15_04.664$firefox 16A-314_2016_02_14_T06_15_04.664/products/pipeline-20160214T061829/html/index.html &
% on-source 30mn, total time: 2016-02-14 05:52:21 UTC -  	2016-02-14 04:41:39 UTC

%%A-array
%%firefox 16B-065_2016_10_20_T15_35_03.021/products/pipeline-20161020T154632/html/index.html  &
%Observation Start 	2016-10-20 11:47:16 UTC
%Observation End 	2016-10-20 15:01:10 UTC
%on-source:  	1:55:12
%
%%firefox 16B-065_2016_10_22_T15_35_04.477/products/pipeline-20161022T154429/html/index.html  &
%Observation Start 	2016-10-22 11:49:15 UTC
%Observation End 	2016-10-22 15:03:12 UTC
%on-source: 	1:55:12
%
%Observation Start 	2016-10-24 11:14:23 UTC
%Observation End 	2016-10-24 14:28:18 UTC
%1:55:16

%B-array

%16A-314_2016_08_25_T14_15_36.478
%Observation Start 	2016-08-25 11:43:48 UTC
%Observation End 	2016-08-25 13:48:24 UTC
%1:02:24
%
%16A-314_2016_08_29_T14_35_00.521
%Observation Start 	2016-08-29 10:36:54 UTC
%Observation End 	2016-08-29 12:41:24 UTC
%1:02:21
%
%
%16A-314_2016_08_30_T15_56_22.120
%Observation Start 	2016-08-30 09:34:59 UTC
%Observation End 	2016-08-30 11:39:33 UTC
%1:02:03
%
%16A-314_2016_08_30_T15_56_23.732
%Observation Start 	2016-08-30 11:40:41 UTC
%Observation End 	2016-08-30 13:45:15 UTC
%1:02:18

\subsection{Instrumental setup}

The new VLA data were acquired in array configurations C, B (project
ID 16A-314), and A (project ID 16B-065), and were executed in a total
of 10 scheduling blocks (SBs, Table\,\ref{table:log}).  The correlator
setup was common to all projects, and covered from 28.976 GHz to
37.024 GHz in 64 spectral windows, each divided into 64 channels, and
with a center frequency of 33.0\,GHz. This corresponds to the Ka band
of the VLA. The bandpass and amplitude calibrators are also listed in
Table\,\ref{table:log}. The phase calibrator was J0559+2353 and common
to all SBs. We typically integrated for 3m18s on target in A and B
array configurations, and for 3m03s in C-configuration, before
switching to the phase calibrator for 1m03s. All datasets were
processed by the VLA pipeline (CASA 4.3.1), and required only a small
amount of posterior flagging to eliminate particularly noisy
combinations of baselines and spectral windows.

The absolute astrometric accuracy of the data may be affected by a
faulty atmospheric delay correction while the A-configuration data
were acquired. This small error was subsequently fixed by the
observatory with a new pipeline processing, and this work is based on
the reprocessed data. Nonetheless, the point source at the stellar
position was offset by $\sim$43\,mas to the North from the nominal
stellar position after correction for proper motion. This offset is
much larger than the positional error inferred from the Gaia
catalogue, and it is comparable to the clean beam in A-configuration,
we therefore assumed that the astrometric calibration of these VLA
data are not reliable, and proceeded to fix the origin of coordinates
to the centroid of an elliptical Gaussian fit to the central point
source.  Interestingly, this choice also improved the centering of the
stellar signal in B-configuration, which should not be appreciably
affected by the faulty atmospheric delays picked up by the
observatory.

%(3.677, -26.508)
%from proper motion to 10-2017:
%Exact WCS Centroid: 5:30:27.5334  25:19:56.6491 
%Linear WCS Centroid -0.000448816727747214 0.0428403369804007

%from CRVAL in ALMA data.
%VLA star is  0.042\arcsec  offset towards: 
%Linear WCS Centroid -9.61167130136801e-06 -6.79329991137362e-06
%Bmax/2 1.68733675290113e-05; Bmin/2 1.43295440731903e-05;
%PA-23.3373667878203deg (South of East);
%
%DATE ALMA DATA: 2015-09-01
%CRVAL in ALMA data:
%mod_out_ALMA.fits> 8.261473314494e+01 2.533240932686e+01 
%currentdRA = 82.6147202983692
%currentdDEC = 25.3324059719444
%

%Exact WCS Centroid: 5:30:27.5332  25:19:56.6439 
%       current RA = 5:30:27.5333  25:19:56.6151

%C-array:
%16A-314
%0137+331=3C48: bandpass+ flux calib
%J0559+2353, phase+pointing
%
%B-array:
%16A-314
%0521+166=3C138 amplitude/bandpass
%J0559+2353 phase 
%
%A-array:
%16B-065
%J0559+2353 phase
%J0521+1638 bandpass
%

\begin{table} 
\caption{VLA observations of MWC\,758. \label{table:log}}
\begin{center}
\begin{tabular}{lllllll}
Array     &  start$^a$        & $\Delta T$$^b$ &  $\Delta T_\mathrm{on}$$^c$   & $\tau_{34}^d$ & fluxcal$^e$    \\  \hline
C         &  02-14 04:41  & 1h10m  & 30m    &  0.029 &  3C48  \\
B         &  08-25 11:44  & 2h04m  & 1h02m  &  0.047 &  3C138  \\
B         &  08-27 11:12  & 2h04m  & 1h02m  &  0.049 &  3C138  \\
B         &  08-27 14:45  & 2h04m  & 1h02m  &  0.049 &  3C138  \\
B         &  08-29 10:37  & 2h04m  & 1h02m  &  0.047 &  3C138  \\
B         &  08-30 09:35  & 2h04m  & 1h02m  &  0.049 &  3C138  \\
B         &  08-30 11:41  & 2h04m  & 1h02m  &  0.048 &  3C138  \\
A         &  10-20 11:47  & 3h14m  & 1h55m  &  0.038 &  3C138  \\
A         &  10-22 11:49  & 3h14m  & 1h55m  &  0.038 &  3C138  \\
A         &  10-24 11:14  & 3h14m  & 1h55m  &  0.036 &  3C138  \\
\end{tabular}
\end{center}
\begin{center}
  $^a$ Start UTC date of each integration, year 2016\\
  $^b$ Total execution time \\
  $^c$ On-source integration\\
  $^d$ Sky optical depth at 34\,GHz\\
  $^e$ Source for bandpass and amplitude calibration
\end{center}
\end{table}

\subsection{Imaging}

%scales=[0,5,10,20] A-array  , cellsize  0.1
%scales=[0,5,10,20] B-array  , cellsize  0.2
%scales=[0,5,10] C-array  , cellsize  0.5

\begin{figure*}
\begin{center}
  \includegraphics[width=\textwidth,height=!]{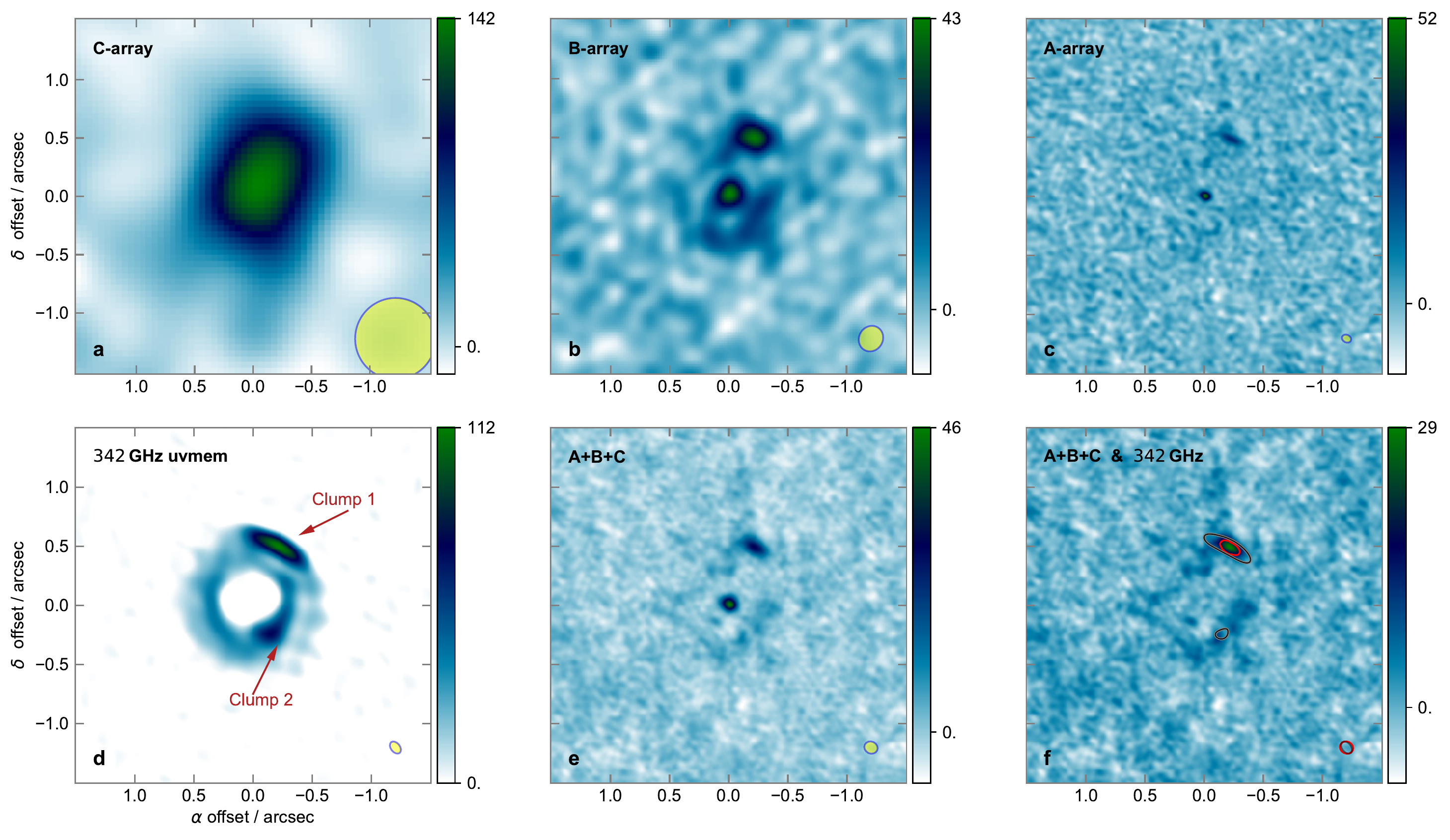}
\end{center}
\caption{Summary of VLA observations and comparison with
  342\,GHz. $x-$ and $y-$ axis show offset in arcsec along RA and DEC,
  with an origin on the star. {\bf a,b,c,e,f:} Clean images obtained
  using Briggs weights (robustness parameter of 1.0) in CASA
  multi-scale {\tt tclean} reconstructions, for each VLA array
  configuration, with the following clean beams and 1\,$\sigma$ noise
  levels: {\bf a}, C-configuration, $(0.70\arcsec\times 0.69\arcsec)$
  beam, along a beam position angle BPA=166\,deg East of North, and
  7.6$\mu$Jy\,beam$^{-1}$ noise; {\bf b}, B-configuration,
  $(0.22\arcsec\times 0.20\arcsec)$ beam, BPA=149\,deg,
  2.9\,$\mu$Jy\,beam$^{-1}$ noise; {\bf c}, A-configuration,
  $(0.08\arcsec\times 0.06\arcsec)$ beam, BPA=64\,deg,
  3.0\,$\mu$Jy\,beam$^{-1}$ noise; {\bf e, f}, combination of all
  three configurations, $(0.115\arcsec\times0.102\arcsec)$ beam,
  BPA=65\,deg, 2.04\,$\mu$Jy\,beam$^{-1}$ noise.  The wedges gives the
  range in specific intensity, in units of $\mu$Jy\,beam$^{-1}$.  {\bf
    d:} Non-parametric model image of the ALMA observations, in units
  of $\mu$Jy\,pixel$^{-1}$, and a pixel size of $0.01\arcsec^2$, with
  an effective resolution $(0.11\arcsec\times0.08\arcsec)$, in the
  direction BPA=38\,deg.  {\bf f:} The grey scale corresponds to the
  same VLA dataset as e), but after subtraction of the central point
  source, and also shown in a single red contour at 60\% peak.  The
  ALMA 342\,GHz image from e) is overlaid in a single black contour,
  also at 60\% peak intensity. Beam ellipses are shown in matching
  colours.  \label{fig:ABC}}
\end{figure*}

A summary of the VLA observations is given in
Fig.\,\ref{fig:ABC}a,b,c,e. These images were obtained with an
application of the multi-scale Clean algorithm
\citep[][]{RauCornwell2011A&A...532A..71R}, using task {\tt tclean}
from the CASA package. As expected, progressively longer baselines
highlight the smaller angular scales in the source. In VLA
B-configuration (Fig.\,\ref{fig:ABC}b), we confirm the detection of
the compact 33\,GHz signal at a position angle (PA) of $\sim$345\,deg
East of North (i.e. $\sim$1\,h on the clock), hereafter Clump\,1,
initially reported by \citet{Marino2015ApJ...813...76M}, and which
also coincides with the peak sub-mm emission \citep[][ALMA Band\,7 at
  337~GHz]{Marino2015ApJ...813...76M}. However, the second clump at
195\,deg (i.e. 5\,h), Clump\,2, which is clearly detected in Band\,7
\citep[][]{Marino2015ApJ...813...76M}, does not appear to coincide
with an equally compact signal at 33\,GHz. In VLA A-configuration
(Fig.\,\ref{fig:ABC}c), we see that most of the disk signal is
resolved out, and Clump\,1 stands out as an arc-like feature, which is
unresolved in the radial direction. The combination of all array
configurations (Fig.\,\ref{fig:ABC}d) recovers extended emission
absent in the longer baselines, at the expense of a coarser clean
beam. Since imaging from a combination of different array
configurations depends on their relative weights, we tested different
combination schemes, and found that the visibility weights as
delivered by the pipeline produced the best results (compared to
reinitialising weights or replacing them by the observed visibility
dispersions).

The peak signal in these Ka maps is the point source at the center,
which likely corresponds to the central star, with a 33\,GHz flux
density of $43.6\pm2.0\,\mu$Jy as given by an elliptical Gaussian fit
using the map shown in Fig.\,~\ref{fig:ABC}e. This flux coincides with
the peak in the map of $46.4\pm2.0\,\mu$Jy\,beam$^{-1}$, within the
errors, as expected for a point source. Indeed, the best fit major and
minor axis for an elliptical fit to the central point source in the
A-configuration map coincide exactly with the beam. In the previous B-configuration
observations at 33\,GHz, from Oct. and Nov. 2013, the stellar flux
amounted to $63\pm5 \mu$Jy. Thus there seems to be a small measure of
stellar variability at 33\,GHz, at 3.6\,$\sigma$. The spectral index
of the point source cannot be determined within the Ka spectral
windows in this new dataset, given the available noise levels, but
\citet{Marino2015ApJ...813...76M} estimate $\alpha^{33{\rm\,GHz}}_{15{\rm\,GHz}} =
0.36\pm0.20$ between Ku and Ka, comparable to the theoretical value
of $0.6$ expected from free-free emission in stellar winds from
early-type stars \citep[][]{WrightBarlow1975MNRAS.170...41W}.

\subsection{Comparison with the sub\,mm continuum} \label{sec:ALMA_VLA}

In the dust trap interpretation for Clump\,1
\citep[][]{Marino2015ApJ...813...76M}, we expect the larger grains to
be progressively more concentrated, until they reach a dimensionless
stopping time (Stokes number) $S_t \sim 1$, when the grains start to
decouple aerodynamically. Since the smaller grains emit more
efficiently at higher frequencies, we compare the VLA data with the
ALMA observations at 342\,GHz recently published by
\citet{Boehler2018ApJ...853..162B}. These ALMA data have coarser
angular resolution than the VLA observations presented here. However,
their very high dynamic range suggests to attempt super-resolution
with a deconvolved model image.

We used the {\tt uvmem} package \citep[][]{Carcamo2018A&C....22...16C}
to fit a model image $\{I^m_i\}$ to the data by minimising the
following objective function:
\begin{equation}
  L = \chi^2 +  \lambda \sum_i   p_i \ln\left(p_i/M\right), \label{eq:L}
\end{equation}
where
\begin{equation}
  \chi^2 = \frac{1}{2} \sum_{k=0}^{N} \omega_k \left\| V^\circ_k - V^m_k\right\|^2.   \label{eq:chi2}
\end{equation}
The free parameters are related to the sky intensity by $p_i = I_i
/\sigma_D$, where $\sigma_D$ is the thermal noise in the
natural-weights dirty map. $M$ is the minimum dimensionless intensity
value; here we set $M =10^{-3}$, and $\lambda = 10^{-3}$. These
choices represent a small amount of image regularization, which
results in slightly less noise compared to the case with $\lambda=0$.
The model image $\{I^m_i\}$ is shown in Fig.\,\ref{fig:ABC}d. The
dirty map of the residual visibilities, in natural weights, are
essentially thermal (with an rms noise of 0.08\,mJy\,beam$^{-1}$).

%For the VLA A+B+C data, image positivity alone provides enough
%regularization, so that we set $\lambda=0$ and $M=0$. The best fit
%model image $\{I^m_i\}$ for the VLA A+B+C data is shown in
%Fig.\,\ref{fig:ABC}g.

The effective angular resolution of the {\tt uvmem} model image can be
estimated by simulating the same $uv-$coverage on a spike, whose flux
is comparable to that of the structures of interest. In the case of
the 342\,GHz data, an elliptical Gaussian fit gives
$(0.11\arcsec\times0.08\arcsec)$, in the direction BPA=38.3\,deg,
which is between 1/3 and 1/2 the natural-weights beam
$(0.33\arcsec\times0.21\arcsec)$.  The {\tt uvmem} effective
resolution is comparable to that of the super-uniform image from
\citet[][ their Fig.\,2]{Boehler2018ApJ...853..162B} of
$0.119\times0.105$ (BPA=66.2\,deg), but it is more elongated. As a
result of the more elongated beam obtained with {\tt uvmem}, Clump 2
seems to vary in radial width, from a broad peak at PA 190\,deg, to a
narrower tail at PA 270\,deg. In turn, the super-uniform restored
image is noisier than the {\tt uvmem} image (probably because uniform
weights do not propagate the measurement accuracies). 

The resolution of the A+B+C image at 33\,GHz is very close to that of
the 342\,GHz model image, as shown by the beam ellipses in
Fig.\,\ref{fig:ABC}f. Even though both images are not comparable on
exactly the same footing, their similar angular resolutions allow a
discussion of trends in the brightest structures. Fig.\,\ref{fig:ABC}f
shows that Clump\,1 is markedly more concentrated at 33\,GHz than at
342\,GHz. In turn, Clump\,2 is almost absent at 33\,GHz, where only
faint and extended signal is seen at these resolutions.  We also see
that Clump\,1 aligns fairly well at both frequencies, using the
default ALMA astrometry, and after the correction of the VLA
astrometry (as described above). Note, however, that the pointing
accuracy of the ALMA data is typically 1/10 of the clean beam, or
$\sim$0.03\arcsec, so  any differences less than 0.1\arcsec are
not significant.

\subsection{Clump\,2 at 33\,GHz and imaging at coarse angular resolution} \label{sec:clump2imaging}

\begin{figure}
\begin{center}
  \includegraphics[width=0.8\columnwidth,height=!]{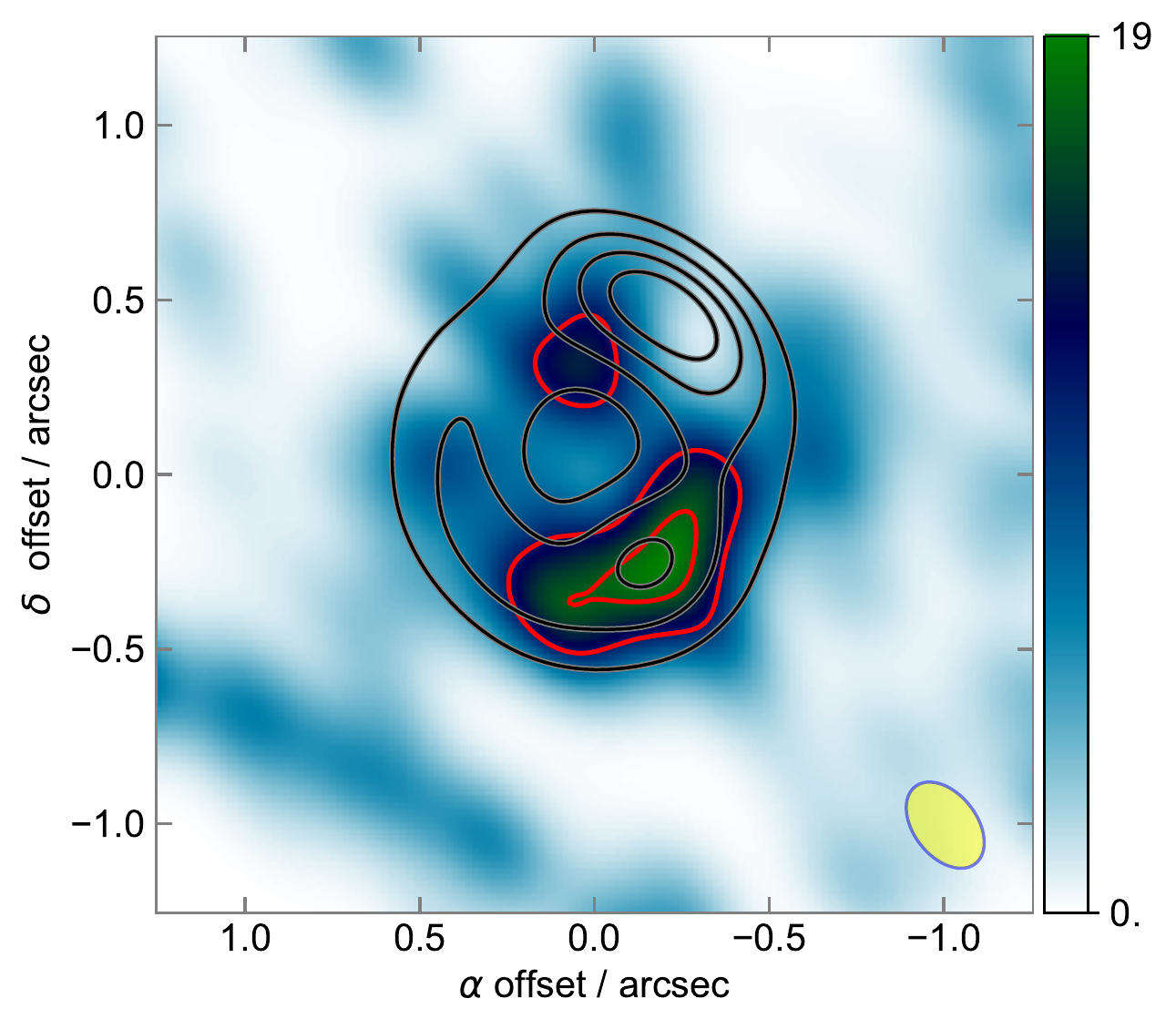}
\end{center}
\caption{Counterpart of Clump\,2 at 33\,GHz. In colour stretch we show
  a restoration of the VLA A+B+C dataset based on a non-parametric
  model image convolved with the clean beam of the ALMA data in
  natural weights, $(0.33\arcsec\times0.21\arcsec)$ along BPA=40\,deg,
  after subtraction of the star and Clump\,1. The red contours follow
  the VLA signal at [0.6, 0.9] times the peak, which for a (thermal)
  noise level of $\sigma=1.6\,\mu$Jy\,beam$^{-1}$ corresponds to
  $[7.2\sigma, 10\sigma]$. For a higher noise level of
  $5\mu$Jy\,beam$^{-1}$ (the peak signal outside the disk), the VLA
  levels would correspond to $[2.3\sigma, 3.5\sigma]$.  Wedge units
  are $\mu$Jy\,beam$^{-1}$. The same restoration for the ALMA 342\,GHz
  data from Fig.\,\ref{fig:ABC}d, but without subtraction of Clump\,1,
  is shown in black and grey contours at [0.2,0.4,0.6,0.8] times the
  peak.  \label{fig:Clump2}}
\end{figure}

The absence of a 33\,GHz counterpart to Clump\,2 in A-configuration
resolutions, while it appears to be detected in B-configuration
(Fig.\,\ref{fig:ABC}b), suggests that Clump\,2 may be more extended
than Clump\,1 at 33\,GHz.  In order to quantify the inter-Clump
spectral trends discussed in Sec.~\ref{sec:contrastC1C2}, we combined
the multi-configuration A+B+C data into a single non-parametric model,
$I^m_{33{\rm\,GHz}}$, without regularisation except for image
positivity (i.e. we minimised $\chi^2$ in Eq.\,\ref{eq:chi2}), as the
inclusion of an entropy term (so with $\lambda >0$ in Eq.\,\ref{eq:L})
eliminates the fainter signal from the model.  We made sure that the
dirty maps of the residual visibilities, in natural weights, were
indeed thermal for each configuration independently.  We then
proceeded to subtract the star and Clump\,1 using elliptical
Gaussians, and degraded this model image to the natural-weights beam
of the 342\,GHz data. Fig.\,\ref{fig:Clump2} compares this coarse
A+B+C 33\,GHz image, $I^c_{33{\rm\,GHz}}$, against
$I^c_{342{\rm\,GHz}}$, the model image at 342\,GHz from
Fig.\,\ref{fig:ABC}d also smoothed by the same beam.

The coarse map $I^c_{33{\rm\,GHz}}$ in Fig.\,\ref{fig:Clump2}, reveals an
intriguing signal {\em inside} the sub\,mm ring, at a PA of
$\sim$5\,deg. Its peak intensity is $14\pm5\mu$\,Jy\,beam$^{-1}$,
which constitutes a tentative detection, at just about 
3\,$\sigma$.

%The nature of this signal will make the object of a forthcoming
%article.

%As a point of comparison, we also show in Fig.\,\ref{fig:cavity}a the
%{\tt tclean} restoration\footnote{image restoration is achieved by
%  addition of the dirty map of residuals visibilities to a convolution
%  of the model with the clean beam} with the same coarse beam.

%Since the {\tt uvmem} $\chi^2$ restoration recovers more extended
%emission than {\tt clean} (compare Figs.\,\ref{fig:ABC}d and g), and
%since it is easier to subtract the brightest component in the
%deconvolved image, $I^m_{33}$, we base our discussion on $I^c_{33}$
%rather than Fig.\,\ref{fig:cavity}a.

%Bmax/2 0.0579669470623286; Bmin/2 0.038567442164739;
%PA-51.682370436407deg (South of East);

%Chi2:
%Bmax/2 0.0733631079386031; Bmin/2 0.0491577873286079;
%PA-55.031123341617deg (South of East);

% {\bf b:} restored image using natural
%  weights, in units of mJy\,beam$^{-1}$, with a beam of
%  $(0.33\arcsec\times0.21\arcsec)$ along BPA=39.6\,deg. {\bf c:} dirty
%  map of the visibility residuals, in mJy\,beam$^{-1}$, and with the
%  same beam as in b).
%
%~/common/ppdisks/MWC758/spinning_dust/map_Clump2.py

%%map_VLA_ALMA.py
%\begin{figure}
%\begin{center}
%  \includegraphics[width=\columnwidth,height=!]{./figs/fig_VLA_ALMA.pdf}
%\end{center}
%\caption{ Overlay of the ALMA 342\,GHz image (Fig.\,\ref{fig:ALMAmem}a),
%  in a single blue contour at 60\% peak intensity, on the VLA restored
%  imaged from Fig.\,\ref{fig:ABC}d, also shown in a single red
%  contour at 60\% peak. Beam ellipses are shown in matching
%  colours. \label{fig:VLA_ALMA}}
%\end{figure}
%

\subsection{Radial extent of  Clump\,1 and aspect ratio} \label{sec:radialwidth}

\begin{figure}
\begin{center}
  \includegraphics[width=\columnwidth,height=!]{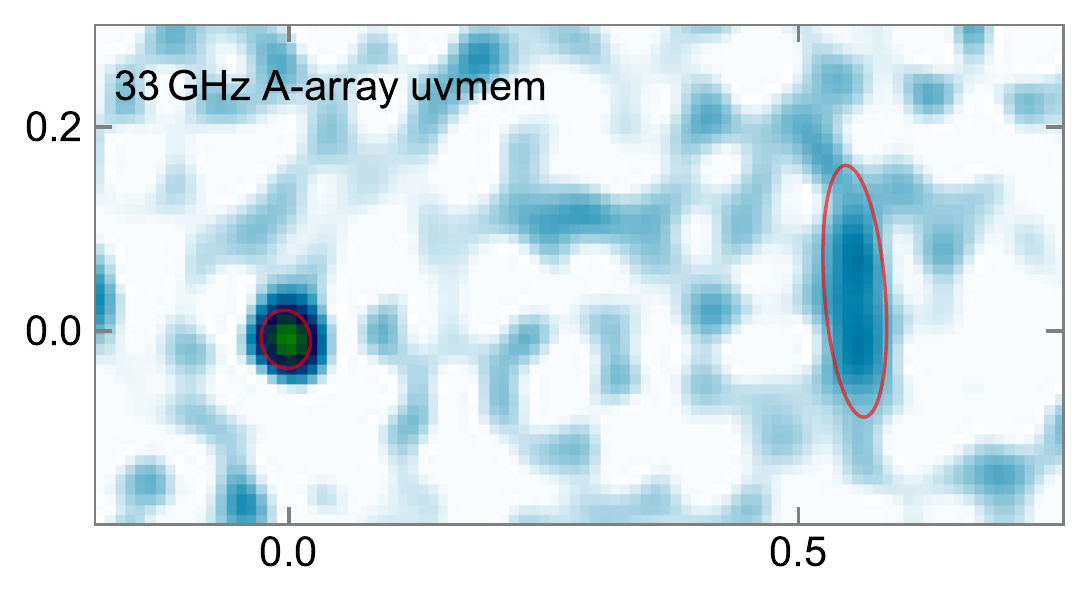}
\end{center}
\caption{Zoom on the star and Clump~1 in a deconvolution of the VLA
  A-configuration dataset, after deprojection to account for the disk
  inclination. The field has been rotated so that  the disk major axis
  lies in the $y$ direction. The red ellipses correspond to elliptical
  Gaussian fits to both the star and Clump~1.  $x-$ and $y-$axis
  indicate offset in arcsec.  \label{fig:mem_A}}
\end{figure}

Interestingly, Clump\,1 appears to be unresolved in the radial
direction, even in A-configuration. At the time of writing no facility
exists that could provide a finer angular resolution at 33~GHz than
the VLA in A-configuration, so we have recourse to deconvolution of
the A-configuration dataset by itself. We use only image positivity
for regularization, i.e. with $\lambda=0$ in Eq.\,\ref{eq:L} (same as
Sec.\,\ref{sec:clump2imaging}), as this choice optimizes angular
resolution (at the expense of a noisier model image).  The discussion
on the physical processes in Clump\,1 depends on its intrinsic width
and aspect ratio $\chi$, so we stretched this deconvolved image to
compensate for the projection at finite inclination.  The resulting
model image, shown in Fig.~\ref{fig:mem_A}, is slightly noisier than
the model for the combined A+B+C dataset, but has a finer effective
angular resolution: the elliptical Gaussian fit to the star is
$(0.057\arcsec \times 0.048\arcsec)$ (a simulation on a spike gave a
similar result). Another Gaussian fit to Clump\,1 gives $(0.248
\arcsec \times 0.060\arcsec)$, with a major axis lying within 7~deg of
the elongation for the stellar signal. Since the orientation of
Clump~1 is approximately parallel to the beam major axis, we limit its
aspect ratio $\chi$ by subtracting the beam in quadrature, after which
the ellipsoidal fit to Clump\,1 is $(0.241\arcsec\pm0.006\arcsec
\times 0.036 \arcsec\pm0.004\arcsec)$, where the uncertainties
correspond to the typical deviations from the Gaussian profile, and do
not include systematics. The aspect ratio of Clump~1 in these
deconvolved maps would thus be $\chi = 5.94 \pm 0.11$. However, we
have not considered the systematics in the error budget, and since
Clump~1 is but marginally resolved in the radial direction, in these
optimistic errors, we report a $3\sigma$ lower limit of $\chi > 5.6$.

%A 1-D Gaussian fit through the arc in a polar map of the deconvolved
%model image from Fig.\,\ref{fig:ABC}g gives a FWHM of $55\pm4$\,mas
%. After correction for the effective angular
%resolution of the {\tt uvmem}-$\chi^2$ model image, of $\sim$30mas, we
%estimate a radial width $\delta r \sim 46\pm4$\,mas FWHM, or $6.9\pm
%0.6$\,au. This deconvolved radial extent can be compared to the arc
%azimuthal length in the {\tt uvmem} image,
%$s^\circ_\phi(33)=28.0\pm1$\,au, after subtraction of the effective
%angular resolution, to yield a vortex aspect ratio of $\chi \approx
%4.03 \pm 0.09$, as seen in the VLA data, which is minimally affected
%by the finite optical depth compared to the ALMA data.

\section{Discussion} \label{sec:discussion}

%Clump\,1 is markedly more compact at 33\,GHz than at 342\,GHz, also as
%expected from azimuthal dust trapping.

\subsection{Azimuthal dust trapping in Clump\,1: spectral trends} \label{sec:trends}

\begin{figure}
  \begin{center}
    \includegraphics[width=\columnwidth,height=!]{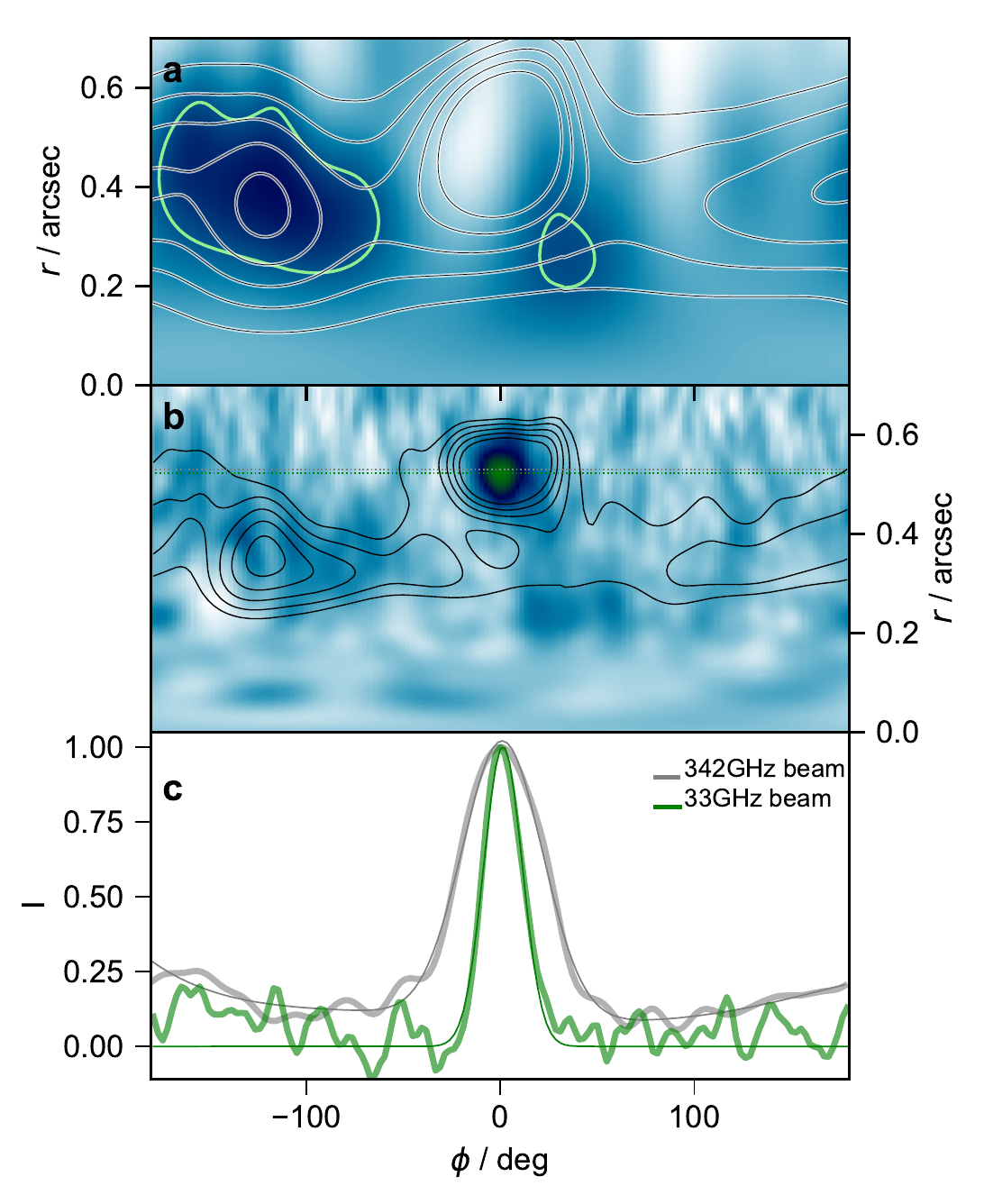}
  \end{center}
\caption{Comparison of the azimuthal profiles for Clump 1 at 342\,GHz
  and 33\,GHz. {\bf a:} Polar map of the VLA A+B+C image from
  Fig.\,\ref{fig:Clump2}, so after subtraction of the star and
  Clump\,1, and with a single light green contour at 0.7 times the
  peak (this level corresponds to 8.4$\sigma$, for a 1$\sigma$ noise
  of $1.6\,10^{-6}$\,$\mu$Jy\,beam$^{-1}$).  The black and grey
  contours correspond to the ALMA image from Fig.\,\ref{fig:Clump2},
  i.e. in the natural-weights beam, with levels at 0.2, 0.3, 0.4, 0.5,
  0.6 times the peak.  The origin of coordinates for the polar
  expansion is offset from the star by 60\,mas towards 28\,deg, so
  that the star would be found at $\sim$125\,deg,0.06\arcsec in polar
  coordinates. {\bf b:} Polar map of the VLA A+B+C image from
  Fig.\,\ref{fig:ABC}e. The contours correspond to the deconvolved
  ALMA image from Fig.\,\ref{fig:ABC}d, with levels at 0.2, 0.3, 0.4,
  0.5, 0.6 times the peak.  The origin of coordinates is offset from
  the star by 60\,mas towards 28\,deg. {\bf c:} Intensity profiles, in
  thick lines, extracted at constant radii along the thin dotted lines
  in b). The 33\,GHz profile is shown in green, while the 342\,GHz
  profile is shown in grey. The thin black lines correspond to
  Gaussian fits with polynomial baselines. The legends indicate the
  size of the beam major axis at 33\,GHz and
  342\,GHz.  \label{fig:polarmap}}
\end{figure}

At 33\,GHz the emission from the disk is essentially confined to
Clump\,1, with a barely detectable counterpart from the more extended
emission at 342\,GHz, as predicted by
\citet{Marino2015ApJ...813...76M} using the Lyra-Lin steady-state
trapping prescriptions \citep[][hereafter
  LL13]{LyraLin2013ApJ...775...17L}. Thanks to the new VLA
observations, we can now place constraints on the azimuthal extent of
Clump\,1. We expanded in polar coordinates the images shown in
Fig.\,\ref{fig:ABC}e and Fig.\,\ref{fig:ABC}e. The 342\,GHz ring
appeared to be significantly off-centre, so  we manually searched
for an adequate origin, in which the ring is the closest match to a
projected circle \citep[using the orientation parameters from][i.e. an
  inclination of 21\,deg along a disk PA of
  62\,deg]{Boehler2018ApJ...853..162B}. Placing the origin offset by
60\,mas from the star, towards 28\,deg East of North, produced the
polar maps shown in Fig.\,\ref{fig:polarmap}. Clump\,1 is essentially
an arc, whose radius is 0.530\arcsec at 342\,GHz and 0.522\arcsec at
33\,GHz. The peaks at both frequencies are remarkably coincident,
within the 30\,mas pointing uncertainty of the ALMA data. We note that
this pointing accuracy is 1\,$\sigma$, so  with these ALMA data we
cannot constrain the small offsets between the cm and mm-grains 
suggested by \citet{BZ2016MNRAS.458.3927B}.

%~/common/ppdisks/MWC758/genfigs/
%map_polar_wcoarseALMA.py

%{\bf c:} Profiles from Clump\,1 extracted on the RT predictions from
%\citet{Marino2015ApJ...813...76M}, also with Gaussian fits.

A Gaussian fit to the intensity profiles extracted at constant radii
gives a fairly good representation of the azimuthal extent of Clump\,1
(Fig.\,\ref{fig:polarmap}c). For the 342\,GHz {\tt uvmem} model image,
we obtain a FWHM of $w_\phi(342{\rm\,GHz}) = 47.5\pm1$\,deg (we also
included a polynomial baseline). After subtraction of the effective
angular resolution, of 12\,deg in azimuthal angle, the length of the
arc is $s_\phi(342{\rm\,GHz}) = 0.42 \pm 0.01$\,arcsec.  For the A+B+C
deconvolved {\tt uvmem} model, so at 33\,GHz, we obtain
$w_\phi(33{\rm\,GHz})=20.6\pm0.7$\,deg (this width is slightly
different from that reported in Sec.\,\ref{sec:radialwidth} because
here we are using all three array configurations in the {\tt uvmem}
model to improve dynamic range).  After correction for the effective
resolution (of 6.6\,deg in azimuth), the arc length is
$s_\phi(33{\rm\,GHz}) = 0.18 \pm 0.006$\,arcsec. Clump\,1 is broader
at 342\,GHz compared to 33\,GHz by a factor $Q_\phi =
s_\phi(342{\rm\,GHz})/s_\phi(33{\rm\,GHz}) = 2.26\pm0.04$, when using
the {\tt uvmem} deconvolution at both frequencies.

%The uncertainties on the widths are estimated based on the deviations
%from the Gaussian fit (the maximum deviation is taken as 3$\sigma$).

%We may compare the extent of the azimuthal dust trapping against the
%predictions from \citet{Marino2015ApJ...813...76M}, after scaling for
%the new GAIA distance. The Gaussian fits to the synthetic profiles in
%Fig.\,\ref{fig:polarmap}c show that Clump\,1 should be 1.25 times
%broader at 342\,GHz, which is significantly less that in the
%observations. \textcolor{red}{***Expand on the origin of this
%  discrepancy and if it can be alleviated by tweaking the steady state
%  parameters***}

%corresponds to
%an equivalent black-body temperature of 20.4\,K, while the
%stellocentric distance of Clump\,1 of 80\,au corresponds to a
%temperature of 35\,K in the radiative transfer (RT) model from

%\,/common/ppdisks/MWC758/trapping$ perl optical_depth.pl

We caution that the trends for a more compact dust trap at 33\,GHz are
affected by the finite continuum optical depth at 342\,GHz, as in
HD\,142527
\citep[][]{Casassus2015ApJ...812..126C}. \citet{Boehler2018ApJ...853..162B}
quote a maximum optical depth of 0.7 in the continuum. Indeed, with
their midplane temperature of 35\,K at 80\,au (so under Clump\,1), the
peak intensity at 342\,GHz in our deconvolved image, of
112\,$\mu$Jy\,pix$^{-1}$ in Fig.\,\ref{fig:ABC}f, gives an optical
depth of $\tau(342{\rm\,GHz})=0.66$. At 33\,GHz, the peak in Clump\,1,
of 0.43\,$\mu$Jy\,pix$^{-1}$, gives $\tau(33{\rm\,GHz})=0.17$. The
corresponding grain emissivity index is $\beta = 0.57$, a value
typical of large grains. However, the arguments below
(Sec.\,\ref{sec:continuousgrains}) suggest that the midplane gas
temperature could be lower, and perhaps closer to 20\,K. If so, the
sub\,mm emission would be completely optically thick.  A higher-frequency
image is required to estimate the continuum temperature and convert
these spectral trends in terms of variations of the dust emissivity
spectral index.

\subsection{Dust trapping predictions}

\subsubsection{Two-grain-size trapping model  in  Clump\,1} \label{sec:2sizes}

Here we compare the observed size ratio $Q_\phi =
s_\phi(342{\rm\,GHz})/s_\phi(33{\rm\,GHz})$ with the steady state
vortex dust trapping prescriptions from LL13. We assume that each
frequency $\nu_i$ can be approximately ascribed to a single effective
gran radius $a_i$, as is the case for compact grains with a narrow
peak in their absorption opacity as a function of grain size at $a_i =
\lambda_i /(2 \pi)$ \citep[e.g.][their Fig.~11 also shows that fluffy
  grains, with a volume filling factor $f<1$, lack a narrow
  peak]{Kataoka2014A&A...568A..42K}. The optical depth at ALMA
frequency is likely to be fairly high (Sec.~\ref{sec:trends}), but for
simplicity we assume optically thin emission at both frequencies, so
that the spatial distribution of grains can be inferred directly from
the continuum images.

The aerodynamic coupling of gas and dust is described by the
dimensionless stopping time, or Stokes number:
\begin{eqnarray}
  {\rm St}(a) & = & \sqrt{\frac{\pi}{8}}\frac{a}{H} \frac{\rho_\bullet}{\rho_g},  \\
 & = &  \frac{\pi}{2}  \frac{a \rho_\bullet}{\Sigma_g} \label{eq:St}
\end{eqnarray}
which depends on the internal density of the solids $\rho_\bullet$, on
the gas temperature via the disk scale height $H$, and on the gas
volume density $\rho_g$, or on the total gas surface density
$\Sigma_g$. In the limit of small Stokes numbers $\rm{St} \ll 1$ (to
be checked {\em a posteriori}), the vortex scale length is
\begin{equation}
H_V(a) = \frac{H}{f(\chi)} \sqrt{\frac{1}{(S(a)+1)}}, \label{eq:HV}
\end{equation}
where $H$ is the disk scale height, $S(a) = {\rm St}(a)/\delta$, and
where $\delta$ is a diffusion parameter taken to be equal to $\alpha$
\citep[the gas viscosity parameter][]{1973A&A....24..337S}. $f(\chi)$
is a function of the vortex aspect ratio $\chi$, % (Eq.\,35 in LL13
\begin{equation}
  f(\chi)=\sqrt{2 \omega_V \chi - (2 \omega_V^2 + 3) / (1+\chi^2)}.
\end{equation}
LL13 consider two vortex solutions, GNG and Kida, that determine the
vorticity $\omega_V$ and hence the function $f(\chi)$: $\omega_V = 3 /
(2 (\chi -1))$ for Kida, and $\omega_V = \sqrt{3 /(\chi^2 -1 )}$ for
GNG.

The azimuthal extension of the dust distribution for size $a$ is
$s_\phi(a) \sim \chi H_V(a)$. As shown in LL13, $\chi$ does not depend
on grain size. The steady state prediction is
\begin{equation}
  Q_{\phi}  =  \frac{s_{\phi}(a_{342{\rm\,GHz}})}{s_{\phi}(a_{33{\rm\,GHz}})}
  =   \sqrt{\frac{S(a_{33{\rm\,GHz}})+1}{S(a_{342{\rm\,GHz}})+1}},
\end{equation}
where $a_{342{\rm\,GHz}}$ and $a_{33{\rm\,GHz}}$ are average grain
sizes accounting for the continuum emission at each frequency. We
assume that these sizes are proportional to the central wavelengths of
each multifrequency dataset, so $\zeta = a_{33{\rm\,GHz}} /
a_{342{\rm\,GHz}} \approx 10.06$.

%\footnote{these sizes are often taken to be roughly given by $ 2 \pi a
%  = \lambda$ for compact Mie spheres}

Because $S \propto {\rm St} \propto (a \rho_\bullet)$, we have that
$S(a_{\rm 33{\rm\,GHz}}) = \zeta \, S(a_{342{\rm\,GHz}})$, if
$\rho_\bullet(a)$ is constant, and thus
\begin{equation}
  Q_\phi =\sqrt{\frac{\zeta S(a_{342{\rm\,GHz}})+1}{S(a_{342{\rm\,GHz}})+1}}. \label{eq:QS}
\end{equation}
Given the observed value of $Q_\phi=2.26\pm0.04$, we have $S(a_{342{\rm\,GHz}}) =
0.83 \pm 0.16$, and $S(a_{33{\rm\,GHz}}) = 8.34\pm 1.58$.

% $a_{\rm 342} = 0.014$\,cm, and

Since this analysis is restricted to compact grains, the effective
grain size corresponding to 33~GHz is $a_{\rm 33{\rm\,GHz}} = 0.14$\,cm, and the
typical internal density is $\sim$1--3\,g\,cm$^{-3}$ (corresponding to
full spheres with an interstellar medium (ISM) mix of water ice,
graphite and astronomical silicates). We thus obtain from
Eq.~\ref{eq:St} that

%sigma_gas_2sizemodel.py
%alpha_Sigma_g 0.0790436041529 +- 0.014974973479
%buho12:08:47~/common/ppdisks/MWC758/trapping$ python sigma_gas_2sizemodel.py
%alpha_Sigma_g 0.026347868051 +- 0.00499165782634

\begin{multline}
  \alpha \Sigma_g = \frac{\pi}{2} \frac{a_{\rm 33{\rm\,GHz}} \rho_\bullet}{S(a_{\rm 33{\rm\,GHz}})} \approx
  (2.6\pm0.5)\times10^{-2}\,{\rm g\,cm}^{-2}\\  ~\rm{to}~  (7.9\pm1.5 )\times10^{-2}\,{\rm g\,cm}^{-2}.
\end{multline}
The turbulent velocities for vortices from
\citet{LyraKlahr2011A&A...527A.138L}, with Mach numbers of $M_s \sim
0.05$, correspond to $\alpha = M_s^2 \sim 2.5\times10^{-3}$ (which
gives ${\rm St}(a_{342{\rm\,GHz}})= 2.1\times10^{-3}$ and ${\rm
  St}(a_{33{\rm\,GHz}})= 2.1\times10^{-2}$).  We thus expect that
$\Sigma_g \sim 10.5 \pm 2.0\,$g~cm$^{-2}$ to $31.6 \pm
6.0\,$g~cm$^{-2}$, which is at least a factor of 5 larger than the
peak gas surface density inferred from the CO isotopologues by
\citet{Boehler2018ApJ...853..162B}, of 1.8\,g\,cm$^{-2}$.

We reach the conclusion that to explain the observed spectral trends
with compact grains of two representative sizes, and under the
optically thin approximation, either the turbulence levels or the gas
surface density are larger than expected by a factor $\sim 3$ to
10. Returning to the caveat on optical depths, we note that a lower
observed value of $Q_\phi$, as would result after correction for the
optical depth at ALMA frequencies, would further emphasize these
discrepancies.

\subsubsection{Continuous grain size population} \label{sec:continuousgrains}

The two-size model shows that the Lyra-Lin prescriptions require a
more massive disk, or higher levels of turbulence, for hard spheres
and optically thin emission. Here we consider a model that relaxes
these constraints with the incorporation of finite optical depths and
a distribution of grain sizes, including also a grain volume filling
factor $f$. We can discard dust models based on fractal aggregates, at
least with dimension 2, as then the Stokes numbers would be
independent of grain size, and any spectral trend would be solely
accounted for by optical depth effects without segregation of grain
sizes. We thus restrict to a constant filling factor $f$, and estimate
the physical conditions required to account for the spectral trends of
Clump\,1 in terms of the Lyra-Lin dust trapping prescriptions, using
the opacity laws from \citet{Kataoka2014A&A...568A..42K}.

% a continuous grain population along with

The observed constraints are:
\begin{itemize}
\item the peak intensities,
  \begin{eqnarray}
    \ln(I_{342{\rm\,GHz}}/\mathrm{Jy\,sr}^{-1})=\ln\left(4.75\times10^{10}\right) \pm 0.1, \hfill \\
    \ln(I_{33{\rm\,GHz}}/\mathrm{Jy\,sr}^{-1})=\ln\left(1.84\times10^8\right)\pm 0.05, \hfill 
  \end{eqnarray}
  where the uncertainty at 342\,GHz reflects the absolute calibration uncertainty,
\item the deconvolved arc azimuthal widths, $w_\phi(342{\rm\,GHz}) =
  45.9\pm1$\,deg and $w_\phi(33{\rm\,GHz}) = 20.3\pm0.7$\,deg, at a
  radius of $83.5$\,au from the ring centroid, and
\item the inverse contrasts at each frequency, given by
  \begin{eqnarray}
    \mathrm{min}(I_{342{\rm\,GHz}})/\mathrm{max}(I_{342{\rm\,GHz}}) & = &0.089 \pm 0.027,  \hfill  \\
    \mathrm{min}(I_{33{\rm\,GHz}})/\mathrm{max}(I_{33{\rm\,GHz}})& < & 0.198, \hfill   
  \end{eqnarray}
  where the inverse contrast at 33\,GHz is a $3\,\sigma$ upper
  limit. The 342\,GHz contrast is measured against the minimum in the
  Gaussian and polynomial fit in azimuth (see Sec.\,\ref{sec:trends}),
  so that the details of the polynomial baseline are not involved
  in the optimization.
\end{itemize}
The quoted uncertainties for the arc widths are optimistic as they do
not consider systematics. Our choice of constraining the model with
the Gaussian widths after correction for the effective angular
resolutions allows us to avoid convolution of the models, and thus
report our best fit result in native angular resolutions (see
Fig.\,\ref{fig:modelprofiles} below). We note that the choice of using
the Gaussian widths also minimizes the impact of the different $uv$
coverages, as even if the two maps at 33\,GHz and 342\,GHz have a
similar angular resolution, their exact $uv$ coverage is different.

%\begin{eqnarray}
%  \rho(r, \phi, z) = \rho_b(r,z) \left[ 1+ (c-1) \rho_V(r,\phi,z, S(a)) \right], \,\text{with} \\
%  \rho_V = \exp\left[ -\frac{(r-r_\circ)^2}{2 H_V^2}
%    - \frac{(r_\circ \phi)^2} {2H_V^2\chi^2} \right], \,\text{and}\\ \nonumber
%  c = \exp\left[  f^2(\chi) (S(a)+1) / (2 \chi^2 \omega_V^2)  \right]. \nonumber
%\end{eqnarray}

The dust mass surface density fields per unit grain size is given by
\citep[][their Eq.\,11, which we reproduce here because of a
  typographical error]{Marino2015ApJ...813...76M},
\begin{multline}
  \Sigma(r, \phi, a)  =   \Sigma_\circ(a) \Lambda(r,\phi,a), ~ \text{with}\\
  \Lambda(r,\phi,a) = \left( 1+ (c-1) \exp\left[ -\frac{(r-r_\circ)^2}{2 H_V^2} 
    - \frac{(r_\circ \phi)^2} {2H_V^2\chi^2} \right] \right)  ,  \,\text{and} \\
%  c   =   \exp\left[  r^2_S  f^2(\chi) (S(a)+1) / (2 \chi^2 \omega_V^2)   \right].  \hfill\\
   c   =   \exp\left[  \frac{r^2_S H^2}{\chi^2\omega_V^2} \frac{1}{2  H_V^2 }     \right].  \hfill
\end{multline}
The dimensionless parameter $r_S \sim 1 $ is meant to adjust the
boundary of the vortex solution relative to the sonic radius, which
would be $H/(\chi\omega_V)$ along $r$. Here we choose to set $r_S=1$.
The stellocentric radius of the vortex centre is $r_\circ$.  At this
fixed radius $r_\circ$, the surface density of dust with size $a$ is
normalised to the azimuthally averaged gas surface density $\langle
\Sigma_g \rangle$ at $r_\circ$ and gas-to-dust ratio $\epsilon$,
%\begin{multline}
%      \Sigma_a = \frac{ 2 \pi \langle \Sigma_g \rangle y(a)}{\epsilon \int  d\phi \Lambda (r_\circ,\phi,a)}, \,\text{where} \hfill \\
%  y(a) =  \frac{(-q+4) \, a^{-q+3}}{a_\mathrm{max}^{-q+4} -a_\mathrm{min}^{-q+4}}, \hfill
%\end{multline}
\begin{equation}
      \Sigma_\circ(a) = \frac{ 2 \pi \langle \Sigma_g \rangle (4+q)  a^{3+q}}{ \left(a_\mathrm{max}^{4+q} -a_\mathrm{min}^{4+q}\right) \epsilon  \int  d\phi \Lambda (r_\circ,\phi,a)}, 
\end{equation}
for a population of spherical grains whose size distribution is a
power law with index $q < 0$ from $a_\mathrm{min}$ to
$a_\mathrm{max}$. The surface density of the gas background follows
from $a=0$,
\begin{multline}
  \Sigma_g(r, \phi)  =   \hat{\Sigma}_g \Lambda(r,\phi,a=0), ~ \text{and for $r=r_\circ$,}\hfill \\ 
  \hat{\Sigma}_g  = \frac{ 2 \pi \langle \Sigma_g \rangle}{\int  d\phi \Lambda (r_\circ,\phi,a=0)}. \hfill
\end{multline}

%\citet{dAlessio2001ApJ...553..321D} and
%\citet{Sierra2017ApJ...850..115S} of the

The total optical depth is defined as
\begin{multline}
  \tau(\nu,\phi) =   \tau_\mathrm{abs}(\nu,\phi) +   \tau_\mathrm{sca}(\nu,\phi) \\ \hfill
 = \int da \Sigma(r=r_\circ,\phi,a) (\kappa_\mathrm{abs}(\nu,a) + \kappa_\mathrm{sca}(\nu,a)), \hfill
\end{multline}
where the absorption and scattering mass opacities,
$\kappa_\mathrm{abs}$ and $\kappa_\mathrm{sca}$, are given by the
analytical approximations from
\citet[][]{Kataoka2014A&A...568A..42K}. We assume optical constants
corresponding to a mix of astrosilicates (60\%) and amorphous
carbon(40\%). In addition, we also allow for the grains to be porous,
with filling factor $f$, by modifying the optical constants using the
Garnett rule, as proposed by \citet{Kataoka2014A&A...568A..42K}, which
also leads to an internal density of $\rho_{\bullet} =2.77
f$\,g\,cm$^{-3}$.

%where $\epsilon_\nu = 1- \omega_\nu = 1 -  \kappa_\mathrm{sca} / (\kappa_\mathrm{abs} + \kappa_\mathrm{sca})$. 

The emergent continuum is treated as that of an isothermal and uniform
slab viewed at normal incidence (i.e. with $\mu=\cos(\theta)=1$, in
spherical coordinates). We assume isotropic scattering, and use the
solution proposed by \citet{MiyakeNakagawa1993Icar..106...20M} and
\citet[][their Eqs. 2, 3 and 4]{dAlessio2001ApJ...553..321D} for the
emergent intensity \citep[see
  also][]{Sierra2017ApJ...850..115S}, with  grain albedos given by
\begin{equation}
  \omega(\nu,\phi) = \frac{\tau_\mathrm{sca}(\nu,\phi)}{\tau_\mathrm{sca}(\nu,\phi) + \tau_\mathrm{sca}(\nu,\phi)}.
\end{equation}
Provided with 1-D model profiles
$I^m_{\nu i}(\phi)$ as functions of azimuth for each frequency
$\nu_i$, we subtract a flat baseline and fit a Gaussian to $I^m_{\nu
  i}(\phi)-{\rm min}(I^m_{\nu i}(\phi))$, so as to yield the
observables: the Gaussian widths, the peak intensities and the
contrasts.

%$J^{\rm 2s}_\nu$.  We adapt their
%Eq.~29, with $\mu=1/\sqrt{3}$, to the case for $\mu=1$:
%\begin{multline}
%  I_\nu(\mu=1,\phi) = 2 J^{\rm 2s}_\nu\left(\tau_\nu/\sqrt{3},\phi\right) \\
%  = 2 B_\nu   \frac{ \sqrt{\epsilon_\nu} \left( \exp\left[-\sqrt{3\epsilon_\nu}\tau_\nu(\phi)/\sqrt{3}\right] - 1\right)}{\exp\left[-\sqrt{3\epsilon_\nu}\tau_\nu(\phi)/\sqrt{3}\right] \left(\sqrt{\epsilon_\nu} - 1\right) - \left(\sqrt{\epsilon_\nu} +1 \right) },
%\end{multline}

With only two frequencies, our simplified observational constraints
number a total of 6 independent data points, while there are 8 free
parameters in the model: $\langle \Sigma_g \rangle$, $T$, $f$,
$\alpha$, $\epsilon$, $a_\mathrm{max}$, $q$, and $\chi$. The model is
therefore under-constrained. We nonetheless explored parameter space
in search of suitable combinations with a Markov chain Monte Carlo
ensemble sampler \citep{MCMC2010CAMCS...5...65G}. We used the {\tt
  emcee} package \citep[][]{emcee2013PASP..125..306F}, with flat
priors, and with 6000 iterations and 1500 walkers, burning-in the
first 1500 iterations when the chains no longer followed systematic
drifts. Both of the Kida and GNG vortex solutions resulted in very
similar optimisations, so for simplicity we quote results for Kida
only.

\begin{figure*}
\begin{center}
  \includegraphics[width=\textwidth,height=!]{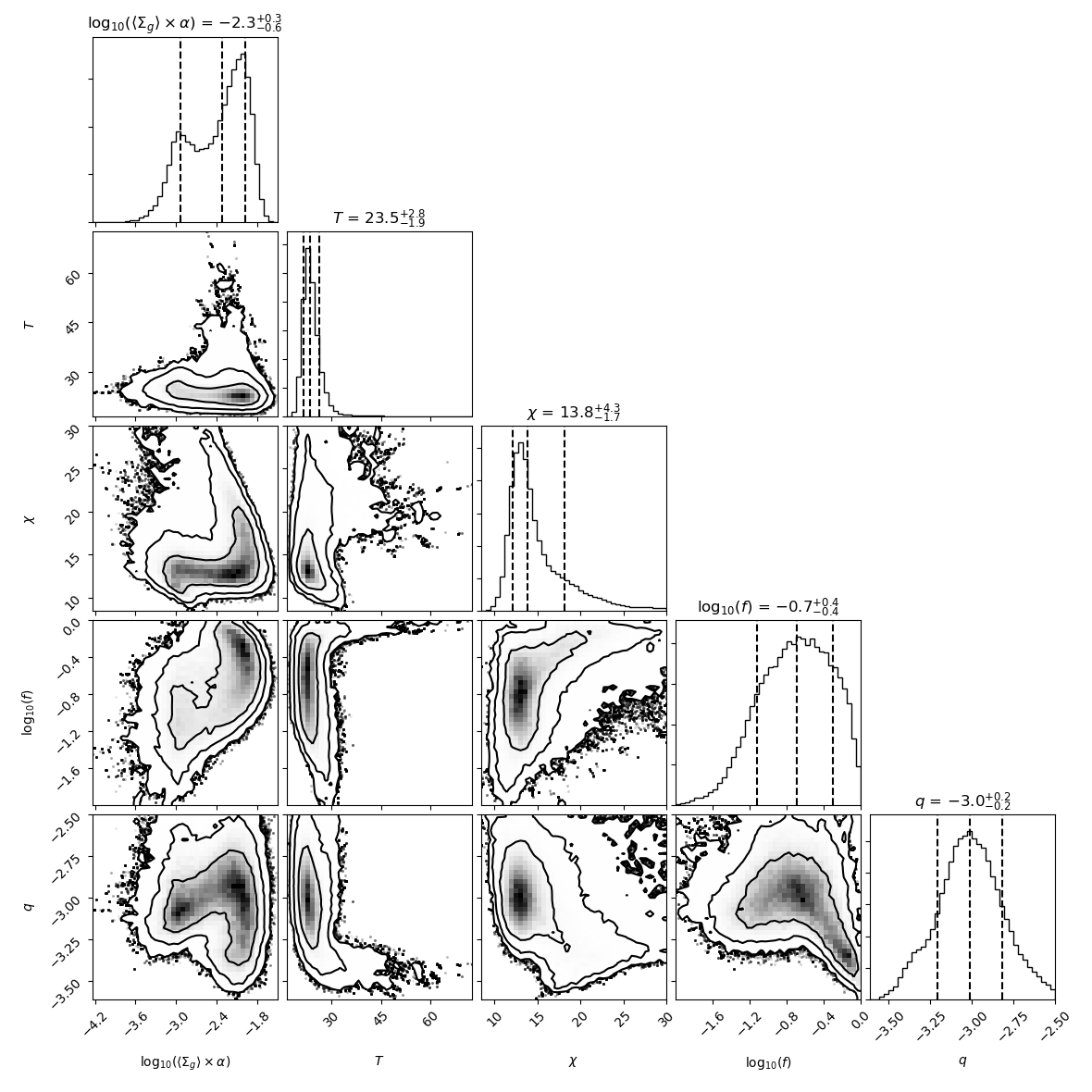}
\end{center}
\caption{Optimization of the isothermal model with continuous grain
  sizes, with all 8 parameters. This corner plot shows the posterior
  probability distributions for 5 selected parameters, the other 3
  being essentially unconstrained. The histograms plot the 1D
  probability density functions of the parameters indicated in titles
  along with their median values and 1$\sigma$ confidence intervals
  (i.e. at 16\% and 84\%), which are also shown by the vertical dashed
  lines.  The contour plots show the marginalized 2D distributions
  (i.e. the 2D projection of the 8D posterior probability
  distribution), for the corresponding pairs of parameters. Contour
  levels are chosen at 0.68,  0.95 and 0.997.
  \label{fig:corner}} 
\end{figure*}

Suitable combinations of parameters occur for a very broad range in
$\langle \Sigma_g \rangle$, which is thus essentially
unconstrained. There is a very tight degeneracy between
$\langle\Sigma_g \rangle$ and $\epsilon$, as the ratio determines the
dust mass and densities, and also with $\alpha$, as the product
determines the Stokes number, suggesting that with such few data
points, fixing $\langle \Sigma_g\rangle$ will be compensated by
$\epsilon$ and $\alpha$. This 8-parameter optimization nonetheless
provided constraints on some key parameters, as shown in
Fig.~\ref{fig:corner}. In particular the allowed temperature values
are narrowly peaked around the median value $T=23.5^{+2.8}{-1.9}$\,K,
where the errors indicate 68\% confidence intervals
(i.e. 1$\sigma$). The vortex aspect ratio is also tightly constrained
to $\chi = 13.8^{+4.3}_{-1.7}$. Filling factors of order unity are
preferred, $\log_{10}(f)= -0.7^{+0.40}_{-0.4}$ (or $f=
0.2^{+0.30}_{-0.13}$).  The median exponent for the grain size
distribution turned out to be fairly flat, $q=-3.0^{+0.2}_{-0.2}$,
although the errors span the whole range of typical values \citep[from
  $-3$ to $-3.5$, also consistent with the slopes found by][ inside
  the vortex]{Sierra2017ApJ...850..115S}.

The model also constrains the product of the average gas surface
density and the turbulence parameter,
\begin{equation}
  \log_{10}\left(\frac{\langle \Sigma_g \rangle \times
    \alpha}{\mathrm{g~cm}^{-2}} \right) = -2.3^{+0.3}_{-0.6}.
\end{equation}
With the limits that we have arbitrarily placed on $\langle \Sigma_g
\rangle /\rm{g\,cm}^{-2}~ \in [0.1,100]$, we can constrain
$\log_{10}(\alpha) < -2.1 $, at 99.7\% confidence (so at 3\,$\sigma$).

%Relaxing the limits requires fitting for the details of the azimuthal
%profiles, rather than just their widths and contrasts, as other
%combinations of parameters produce structured profiles that still bear
%the same widths as the observations.

%
%~/common/ppdisks/MWC758/trapping/optim_wscat/8params_Kida_Isca_emcee2_newdistance/
%python plot_subtriangle.py --> run_results.txt
%
%Sigma_g (43.649268313939885, 23.66195799005547, 24.026384022072573) 37.4828540915
%log10alpha (-3.9012984184441355, 0.44259196317973526, 0.7292086462972578) -3.4443326886
%amax (4.752977196068091, 3.4275323421994663, 3.0117512430749773) 7.31856476066
%epsilon (63.88388014201937, 25.477632813061767, 34.04387479596148) 50.4763936865
%T (23.467229028368152, 2.764794209871603, 1.9232877287304113) 22.6319384728
%chi (13.828009620650068, 4.337686893451213, 1.6873204945824565) 13.8522132701
%f (0.20542929225164913, 0.29932857834902254, 0.12885657874156797) 0.208464736154
%q (-3.013305971207274, 0.1944311344447791, 0.1918106270399771) -2.99287856824
%

%buho19:42:59~/common/ppdisks/MWC758/trapping/optim_wscat/8params_Kida_Isca_emcee2_newdistance$ python plot_subtriangle.py

%ml best fit params:
%Sigma_g 7.55878635961
%log10alpha -2.7403313859
%amax 8.16168080132
%epsilon 10.389234061
%T 22.7025352937
%chi 14.9442122911
%f 0.248069899632
%q -3.07603612998

Fig.\,\ref{fig:modelprofiles} shows the predicted azimuthal intensity
profiles corresponding to the maximum likelihood parameters: $\langle
\Sigma_g \rangle =7.6$\,g\,cm$^{-2}$, $\log_{10}(\alpha)= -2.7$,
$a_{\rm max}=8.2$\,cm, $\epsilon= 10.4$, $T= 22.7$\,K, $\chi = 14.9$,
$f= 0.25$, $q= -3.1$. Models without scattering, so with an albedo
$\omega(\nu,\phi)=0$, yield similar results, i.e. the largest
differences are found for $T=23.6$\,K and $\chi=16.3$, and are all
within the uncertainties.  The model is, unsurprisingly, exactly
coincident with the observations: widths, peak intensities, and the
contrast at 342GHz are reproduced within 1\%. The optical depth
profiles reach optically thick values in the sub-mm, with a peak
absorption optical depth of 2.9 at 342\,GHz. Perhaps more suprising is
the high absorption optical depth in cm-wavelengths, with a peak of
0.3, which, given the observed flux density, reflects the low
temperatures and the small solid angle subtended by the elongated
vortex. Such fairly high optical depths have also been predicted by
\citet{Sierra2017ApJ...850..115S}.

%The corner plot in Fig.~\ref{fig:corner} conveys the distribution and
%correlations, in 5D parameter space, between three selected
%parameters: $\alpha$, $\chi$ and $T$. We see that the distributions
%are well sampled, so that the resulting confidence levels are genuine
%indicators of measurement accuracy. This type of multi-frequency
%analysis thus holds the potential to constrain physical conditions.

%bol14:12:59~/common/ppdisks/MWC758/trapping/optim_wscat/8params_Kida_Isca_emcee2$ python plot_subtriangle.py

%~/common/ppdisks/MWC758/genfigs/profiles_teor.py
\begin{figure}
\begin{center}
  \includegraphics[width=\columnwidth,height=!]{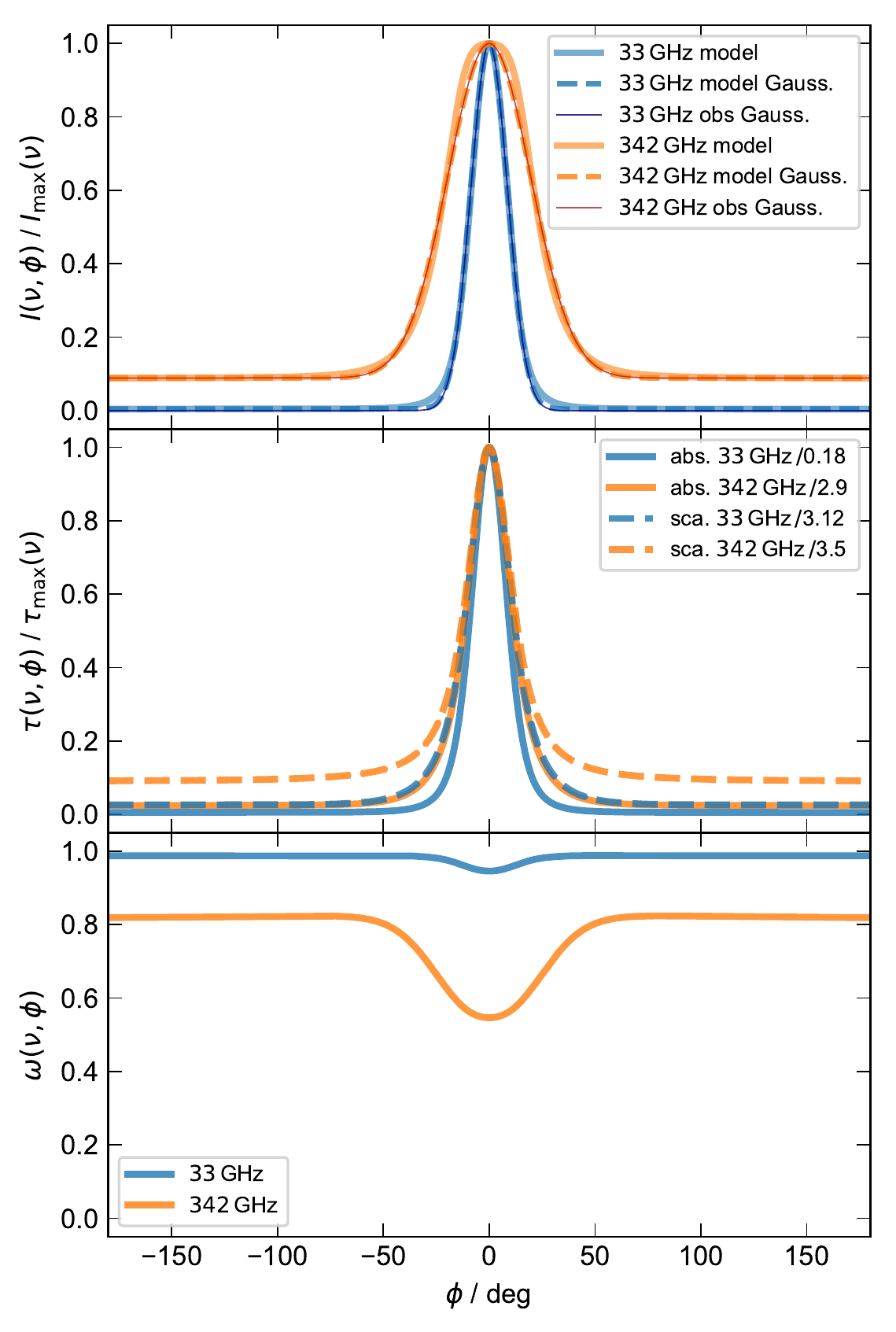}
\end{center}
\caption{Model azimuthal profiles predicted with a continuous grain
  population and the Lyra-Lin trapping prescriptions, and comparison
  with the Gaussian fits to the observed profiles, with flat
  baselines.  {\bf Top:} Intensity profiles. The maximum likelihood
  models at each each frequency are shown in thick solid lines. The
  optimization involves the comparison of a Gaussian fit to the model
  profiles (dashed) against the Gaussian fits to the observed profiles
  corrected for the effective angular resolutions (thin solid lines,
  see text for details). The almost perfect match reflects the reduced
  number of independent data points compared to the free
  parameters. {\bf Middle:} Model optical depth profiles, normalized
  to the peak values indicated in the legend. {\bf Bottom:}
  Corresponding grain albedos at each
  frequency.  \label{fig:modelprofiles}}
\end{figure}

%Compact spheres are preferred, with $f\sim1$.  This results from the
%adopted opacity law: with smaller values of $f\ll 1$, the dust
%emissivity in the mm- to cm-wavelength range is progressively less
%dominated by a well defined characteristic grain size.  Rather, all
%grains smaller than a given size\footnote{this size corresponds to
%  dust grains small enough to be in the Rayleigh regime or still be
%  optically thin at these wavelengths} contribute significantly.
%Thus, the differences in surface brightness between different
%wavelengths becomes less pronounced.

The best fit parameters are all within the expected range, given the
limits, except for the temperature $T=23.5^{+2.8}_{-1.9}$\,K, which is
significantly colder than the 35\,K from the RT predictions in
\citet{Boehler2018ApJ...853..162B}. This difference may be related to
the dust population in the trap being heavily weighted to large grains
compared to that originating the bulk of the sub-mm
emission. Additional multi-frequency data are required to test this
result for low temperatures, which could be due to the simplifications
in this model, or in missing physical processes in the Lyra-Lin
prescriptions (which are isothermal). An inadequate model would result
in deviations from sufficiently detailed observations.

Another interesting result from the optimization of the isothermal
model and steady-state model is the prediction of a very elongated
vortex, $\chi = 13.8^{+4.3}_{-1.7}$. This is in agreement with
hydrodynamical simulations taylored to MWC\,758, which resulted in
$\chi=15\pm2$. In a companion article (Baruteau et al. 2018), we
reproduce the dust emission in Clumps 1 and 2 by means of dust+gas
hydrodynamical simulations, using a 5$M_{\rm jup}$ body at 132\,AU,
which drives Clump\,1, and a 1$M_{\rm jup}$ body at 33\,AU, which
drives Clump\,2. The external companion is also required to account
for the prominent spiral structure \citep[as
  in][]{Dong2015ApJ...809L...5D}. Its mass is limited by recent $L'$
high-contrast data to less than 10\,M$_{\rm jup}$ assuming ``hot
start'' conditions \cite{Reggiani2014ApJ...792L..23R}. The internal
companion is close to the thermal-IR point-source reported by
\cite{Reggiani2014ApJ...792L..23R}, whose $L'$ magnitude is compatible
with a 5 $M_{\mathrm{jup}}$ planet accreting at a rate of
$10^{-6}M_{\mathrm{jup}}\mathrm{yr}^{-1}$.

%Provided with this model for the distribution of optical depths at
%each frequency, we can revisit the two-grain-size trapping model
%(Sec.~\ref{sec:2sizes}). A Gaussian fit to the absorption optical
%depth profiles from Fig.~\ref{fig:modelprofiles} results in $Q_\phi =
%1.37$. The corresponding value for the Stokes number at 33\,GHz is
%$S=1.10$, so that we require ${\rm St}(a_{33})= 2.7\times10^{-3}$ for
%$\alpha \sim 2.5\times10^{-3}$
%\citep{LyraKlahr2011A&A...527A.138L}. The required gas surface density
%would be $\Sigma_g \sim 100$\,g\,cm$^{-2}$. We reach the conclusion
%that a two-grain-size model cannot be used to represent spectral
%trends over a decade in frequency. Eq.~\ref{eq:QS} may however be
%useful to infer the Stokes number from continuum observations at
%adjacent frequencies.

%\begin{figure*}
%\begin{center}
%  \includegraphics[width=0.7\textwidth,height=!]{./figs/triangle.png}
%\end{center}
%\caption{ Search in parameter space for the best fit $\Sigma_g$,
%  $\alpha$, $\epsilon$ and $T$. The blue lines correspond to the
%  maximum likelihood values, while the dashed lines correspond to the
%  $[0.16, 0.5,0.84]$ quantiles (i.e. the median value $\pm 1\sigma$). 
%  \label{fig:triangle}}
%\end{figure*}

\subsection{The contrast between Clump\,1 and Clump\,2} \label{sec:contrastC1C2}

In Figs.\,\ref{fig:ABC}d and f, the peak intensity ratio $R$ between Clump\,1
and Clump\,2 at 342\,GHz is $R_{342{\rm\,GHz}} = 1.51 \pm 0.01$. Yet, at
33\,GHz, and at a similar angular resolution as the 342\,GHz data,
this ratio is $R_{33{\rm\,GHz}}>4.8$ at 3\,$\sigma$, since the peak in Clump\,1
is $29\pm 2.0$\,$\mu$Jy\,beam$^{-1}$ in Fig.\,\ref{fig:ABC}f.

From the smooted image $I^c_{33{\rm\,GHz}}$ shown in
Fig.\,\ref{fig:Clump2}, we see that Clump\,2 is indeed detected at
33\,GHz, at the same location as at 342\,GHz, and with a local peak
intensity of $19\pm 5$\,$\mu$Jy\,beam$^{-1}$, where the 1\,$\sigma$
noise corresponds to the intensity of the spurious features in the
field\footnote{the thermal noise is only 1.6$\mu$Jy\,beam$^{-1}$, but
  imperfections in the phase calibrations result in systematics that
  we chose to include in the noise level, for conservative
  estimates}. Since the peak of $I^c_{33{\rm\,GHz}}$ without
subtraction of Clump\,1 is 51\,$\mu$Jy\,beam$^{-1}$, we have an
inter-clump contrast $R^c_{33{\rm\,GHz}}=2.7\pm0.3$. In the
natural-weights map $I^c_{342{\rm\,GHz}}$, this contrast is
$R^c_{342{\rm\,GHz}}=1.55\pm0.004$.

Some dust trapping may be at work also for Clump\,2, since the 33\,GHz
signal in Fig.\,\ref{fig:Clump2} appears somewhat more lopsided than
at 342\,GHz (the azimuthal contrast at the radius of Clump\,2 is
$\sim$1.5 in $I^c_{342{\rm\,GHz}}$ and $\sim$2.3 in $I^c_{33{\rm\,GHz}}$). However, the
concentration of the cm-grains is not as effective as for Clump\,1. We
see that Clump\,2 is likely more extended than Clump\,1, explaining
its non-detection in A-configuration angular resolutions. We note that even in
the coarse maps, the inter-clump contrast is a factor $\sim$2 greater
at 33\,GHz than at 342\,GHz, suggesting a deficit of cm-grains in
Clump\,2 when compared with Clump\,1.

Since Clump\,2 is found closer to the cavity edge than Clump\,1, it is
possible that the gas background is denser for Clump\,2, such that the
Stokes numbers are reduced. In the parametric model of
\citet{Boehler2018ApJ...853..162B}, the mid-plane gas densities could
be more than twice denser at 60\,au compared to 80\,au, with
correspondingly lower Stokes number for Clump\,2, which would result
in less efficient trapping. Alternatively, the companion paper by
Baruteau et al. (2018) models Clump\,2 as a decaying vortex.

The continuum optical depths under Clump\,2 are markedly lower than
for Clump\,1. Using the midplane temperature at a stellocentric radius
of 50\,au from \citet{Boehler2018ApJ...853..162B}, of 70\,K, we obtain
$\tau(33{\rm\,GHz}) =0.006$ and $\tau(342{\rm\,GHz}) =0.09 $, with an
index $\beta = 1.17$. The non-detection of Clump\,2 in A-configuration
suggests that it is extended, so it is likely that this high $\beta$
index is intrinsic and not just due to the coarse beam.  In the same
coarse beam, the $\beta$ index for Clump\,1 is 0.95, but Clump\,1 is
very compact and a meaningful $\beta$ index can only be obtained in
the finest angular resolutions (for which $\beta = 0.57$, see
Sec.\,\ref{sec:trends}).

%Using the case of compact icy grains in \citet[][their
%  Fig.4]{Testi2014prpl.conf..339T}, the maximum grain size would be
%$\sim$0.2\,cm in Clump\,2, and $\sim$2\,cm in Clump\,1, for the
%$\beta$ indices as obtained from the coarse beam and the finer angular
%resolutions, respectively.

%After smoothing to the resolutions of the deconvolved 342\,GHz image
%and the A+B+C 33\,GHz image, the predictions from the double vortex
%model in \citet{Marino2015ApJ...813...76M} are 0.98 at 342\,GHz and
%0.82 at 33\,GHz, so that Clump\,2 has a higher specific intensity
%(despite a lower flux density). \textcolor{red}{*** This is not a
%  robust prediction, so rather than disprove it, try to fit the new
%  data with the double-vortex model, or else remove ****}.

\section{Conclusions} \label{sec:conclusion}

We presented new VLA observations at 33\,GHz of MWC\,758 that revealed
disk emission concentrated in three regions, which present interesting
differences with reprocessed 342\,GHz ALMA data:
\begin{itemize}
\item Clump\,1, originally detected by
  \citet{Marino2015ApJ...813...76M}, is an arc with a deprojected
  length of $0.24\arcsec\pm0.006\arcsec$, FWHM. Clump\,1 is
  $2.26\pm0.04$ times more compact in azimuth at 33\,GHz than at
  342\,GHz.  Its radial width is $64$\,mas$\pm$4\,mas, and marginally
  resolved in our deconvolved model of the VLA A-configuration data,
  with an effective resolution of $\sim$48\,mas. Correction for the
  finite angular resolution yields linear dimensions of
  $(37\pm0.95\times6.7\pm0.64)\,\text{au}$, for a distance of 160\,pc,
  and where the uncertainties do not account for systematics. The
  resulting aspect ratio is $\chi > 5.3$ (at 3$\sigma$).
\item Clump\,2, which is the second local maximum at 342\,GHz
  \citep{Marino2015ApJ...813...76M} in the outer disk, is also
  detected at 33\,GHz, but is not as compact as Clump\,1. Its intensity
  ratio relative to Clump\,1 is $\sim$2 times less at 33\,GHz.
\end{itemize}

The spectral trends in Clump\,1 are quantitatively consistent with an
isothermal model for dust trapping in an anticyclonic vortex using the
Lyra-Lin prescription from \citet[][]{LyraLin2013ApJ...775...17L}. The
required physical conditions are all consistent with the body of
information on MWC~758. This model makes a robust predictions for the
vortex aspect ratio $\chi = 13.8^{+4.3}_{-1.7}$ and temperature
$T=23.5^{+2.8}_{-1.9}$\,K. An elongated vortex is consistent with
hydrodynamical simulations taylored to MWC\,758 (Baruteau et al. 2018,
{\em submitted}). This result suggests that the very long arc seen in
the ALMA Band\,4 observations reported by
\citet{Cazzoletti2018arXiv180904160C} in HD\,135344B also correspond
to a very elongated vortex.  The large optical depths required by such
low temperatures and narrow aspect ratios, of order $\sim1$ even in
cm-wavelengths, have also been predicted in the generic models of
\citet{Sierra2017ApJ...850..115S}, and could be tested with further
multi-frequency observations of MWC\,758. The model also constrains
the gas surface densities $\Sigma_g$ and the turbulence parameter
through $\log_{10}\left( \alpha \times \Sigma_g / \mathrm{g\,cm}^{-2}
\right) \sim -2.3^{+0.3}_{-0.6}$. For a standard $\alpha \lesssim
10^{-3}$, $\Sigma_g$ is required to be larger by a factor of 3 to 10
compared to that inferred by \citet{Boehler2018ApJ...853..162B} from
the CO isotopologues.

%We note that the large frequency span of the data precludes reaching
%quantitative constraints on physical conditions if using only two
%representative dust sizes.

The low signal from Clump\,2 at 33\,GHz may be due to a factor
$\gtrsim 2$ larger gas density. This would result in less efficient
aerodynamic coupling than in Clump\,1. Alternatively Clump\,2 may
correspond to trapping in a decaying vortex, as suggested in Baruteau
et al. (2018, {\em submitted}).

\section*{Acknowledgements}

We thank the referee for a thorough review and constructive comments.
Financial support was provided by Millennium Nucleus RC130007 (Chilean
Ministry of Economy), and additionally by FONDECYT grants 1171624 and
3160750, and by CONICYT-Gemini grant 32130007. This work used the
Brelka cluster, financed by FONDEQUIP project EQM140101, hosted at
DAS/U. de Chile. The National Radio Astronomy Observatory is a
facility of the National Science Foundation operated under cooperative
agreement by Associated Universities, Inc. This paper makes use of the
following ALMA data: ADS/JAO.ALMA\#2012.1.00725.S. ALMA is a
partnership of ESO (representing its member states), NSF (USA) and
NINS (Japan), together with NRC (Canada), NSC and ASIAA (Taiwan), and
KASI (Republic of Korea), in cooperation with the Republic of
Chile. The Joint ALMA Observatory is operated by ESO, AUI/NRAO and
NAOJ.

% SP acknowledges CONICYT-Gemini grant 32130007.
%MV acknowledges FONDECYT grant 3160750.

%%%%%%%%%%%%%%%%%%%%%%%%%%%%%%%%%%%%%%%%%%%%%%%%%%



\bibliographystyle{mnras}
\bibliography{merged.bib}

\begin{thebibliography}{}
\makeatletter
\relax
\def\mn@urlcharsother{\let\do\@makeother \do\$\do\&\do\#\do\^\do\_\do\%\do\~}
\def\mn@doi{\begingroup\mn@urlcharsother \@ifnextchar [ {\mn@doi@}
  {\mn@doi@[]}}
\def\mn@doi@[#1]#2{\def\@tempa{#1}\ifx\@tempa\@empty \href
  {http://dx.doi.org/#2} {doi:#2}\else \href {http://dx.doi.org/#2} {#1}\fi
  \endgroup}
\def\mn@eprint#1#2{\mn@eprint@#1:#2::\@nil}
\def\mn@eprint@arXiv#1{\href {http://arxiv.org/abs/#1} {{\tt arXiv:#1}}}
\def\mn@eprint@dblp#1{\href {http://dblp.uni-trier.de/rec/bibtex/#1.xml}
  {dblp:#1}}
\def\mn@eprint@#1:#2:#3:#4\@nil{\def\@tempa {#1}\def\@tempb {#2}\def\@tempc
  {#3}\ifx \@tempc \@empty \let \@tempc \@tempb \let \@tempb \@tempa \fi \ifx
  \@tempb \@empty \def\@tempb {arXiv}\fi \@ifundefined
  {mn@eprint@\@tempb}{\@tempb:\@tempc}{\expandafter \expandafter \csname
  mn@eprint@\@tempb\endcsname \expandafter{\@tempc}}}

\bibitem[\protect\citeauthoryear{{Barge} \& {Sommeria}}{{Barge} \&
  {Sommeria}}{1995}]{Barge_Sommeria_1995A&A...295L...1B}
{Barge} P.,  {Sommeria} J.,  1995, \aap, \href
  {http://adsabs.harvard.edu/abs/1995A%26A...295L...1B} {295, L1}

\bibitem[\protect\citeauthoryear{{Baruteau} \& {Zhu}}{{Baruteau} \&
  {Zhu}}{2016}]{BZ2016MNRAS.458.3927B}
{Baruteau} C.,  {Zhu} Z.,  2016, \mn@doi [\mnras] {10.1093/mnras/stv2527},
  \href {http://adsabs.harvard.edu/abs/2016MNRAS.458.3927B} {458, 3927}

\bibitem[\protect\citeauthoryear{{Benisty} et~al.,}{{Benisty}
  et~al.}{2015}]{Benisty2015A&A...578L...6B}
{Benisty} M.,  et~al., 2015, \mn@doi [\aap] {10.1051/0004-6361/201526011},
  \href {http://adsabs.harvard.edu/abs/2015A%26A...578L...6B} {578, L6}

\bibitem[\protect\citeauthoryear{{Biller} et~al.,}{{Biller}
  et~al.}{2012}]{Biller2012}
{Biller} B.,  et~al., 2012, \mn@doi [\apjl] {10.1088/2041-8205/753/2/L38},
  \href {http://adsabs.harvard.edu/abs/2012ApJ...753L..38B} {753, L38}

\bibitem[\protect\citeauthoryear{{Birnstiel}, {Dullemond}  \&
  {Pinilla}}{{Birnstiel} et~al.}{2013}]{Birnstiel2013A&A...550L...8B}
{Birnstiel} T.,  {Dullemond} C.~P.,   {Pinilla} P.,  2013, \mn@doi [\aap]
  {10.1051/0004-6361/201220847}, \href
  {http://adsabs.harvard.edu/abs/2013A%26A...550L...8B} {550, L8}

\bibitem[\protect\citeauthoryear{{Boehler}, {Weaver}, {Isella}, {Ricci},
  {Grady}, {Carpenter}  \& {Perez}}{{Boehler}
  et~al.}{2017}]{Boehler2017ApJ...840...60B}
{Boehler} Y.,  {Weaver} E.,  {Isella} A.,  {Ricci} L.,  {Grady} C.,
  {Carpenter} J.,   {Perez} L.,  2017, \mn@doi [\apj]
  {10.3847/1538-4357/aa696c}, \href
  {http://adsabs.harvard.edu/abs/2017ApJ...840...60B} {840, 60}

\bibitem[\protect\citeauthoryear{{Boehler} et~al.,}{{Boehler}
  et~al.}{2018}]{Boehler2018ApJ...853..162B}
{Boehler} Y.,  et~al., 2018, \mn@doi [\apj] {10.3847/1538-4357/aaa19c}, \href
  {http://adsabs.harvard.edu/abs/2018ApJ...853..162B} {853, 162}

\bibitem[\protect\citeauthoryear{{C{\'a}rcamo}, {Rom{\'a}n}, {Casassus},
  {Moral}  \& {Rannou}}{{C{\'a}rcamo}
  et~al.}{2018}]{Carcamo2018A&C....22...16C}
{C{\'a}rcamo} M.,  {Rom{\'a}n} P.~E.,  {Casassus} S.,  {Moral} V.,   {Rannou}
  F.~R.,  2018, \mn@doi [Astronomy and Computing]
  {10.1016/j.ascom.2017.11.003}, \href
  {http://adsabs.harvard.edu/abs/2018A%26C....22...16C} {22, 16}

\bibitem[\protect\citeauthoryear{{Carrasco-Gonz{\'a}lez}
  et~al.,}{{Carrasco-Gonz{\'a}lez}
  et~al.}{2016}]{CarrascoGonzalez2016ApJ...821L..16C}
{Carrasco-Gonz{\'a}lez} C.,  et~al., 2016, \mn@doi [\apjl]
  {10.3847/2041-8205/821/1/L16}, \href
  {http://adsabs.harvard.edu/abs/2016ApJ...821L..16C} {821, L16}

\bibitem[\protect\citeauthoryear{{Casassus} et~al.,}{{Casassus}
  et~al.}{2013}]{Casassus2013Natur}
{Casassus} S.,  et~al., 2013, \mn@doi [\nat] {10.1038/nature11769}, \href
  {http://adsabs.harvard.edu/abs/2013Natur.493..191C} {493, 191}

\bibitem[\protect\citeauthoryear{{Casassus} et~al.,}{{Casassus}
  et~al.}{2015}]{Casassus2015ApJ...812..126C}
{Casassus} S.,  et~al., 2015, \mn@doi [\apj] {10.1088/0004-637X/812/2/126},
  \href {http://adsabs.harvard.edu/abs/2015ApJ...812..126C} {812, 126}

\bibitem[\protect\citeauthoryear{{Cazzoletti} et~al.,}{{Cazzoletti}
  et~al.}{2018}]{Cazzoletti2018arXiv180904160C}
{Cazzoletti} P.,  et~al., 2018, preprint, \href
  {http://adsabs.harvard.edu/abs/2018arXiv180904160C} {} (\mn@eprint {arXiv}
  {1809.04160})

\bibitem[\protect\citeauthoryear{{Christiaens} et~al.,}{{Christiaens}
  et~al.}{2018}]{Christiaens2018A&A...617A..37C}
{Christiaens} V.,  et~al., 2018, \mn@doi [\aap] {10.1051/0004-6361/201629454},
  \href {http://adsabs.harvard.edu/abs/2018A%26A...617A..37C} {617, A37}

\bibitem[\protect\citeauthoryear{{Cuzzi}, {Hogan}  \& {Shariff}}{{Cuzzi}
  et~al.}{2008}]{Cuzzi2008ApJ...687.1432C}
{Cuzzi} J.~N.,  {Hogan} R.~C.,   {Shariff} K.,  2008, \mn@doi [\apj]
  {10.1086/591239}, \href {http://adsabs.harvard.edu/abs/2008ApJ...687.1432C}
  {687, 1432}

\bibitem[\protect\citeauthoryear{{D'Alessio}, {Calvet}  \&
  {Hartmann}}{{D'Alessio} et~al.}{2001}]{dAlessio2001ApJ...553..321D}
{D'Alessio} P.,  {Calvet} N.,   {Hartmann} L.,  2001, \mn@doi [\apj]
  {10.1086/320655}, \href {http://adsabs.harvard.edu/abs/2001ApJ...553..321D}
  {553, 321}

\bibitem[\protect\citeauthoryear{{Dong}, {Zhu}, {Rafikov}  \& {Stone}}{{Dong}
  et~al.}{2015}]{Dong2015ApJ...809L...5D}
{Dong} R.,  {Zhu} Z.,  {Rafikov} R.~R.,   {Stone} J.~M.,  2015, \mn@doi [\apjl]
  {10.1088/2041-8205/809/1/L5}, \href
  {http://adsabs.harvard.edu/abs/2015ApJ...809L...5D} {809, L5}

\bibitem[\protect\citeauthoryear{{Dong} et~al.,}{{Dong}
  et~al.}{2018}]{Dong2018ApJ...860..124D}
{Dong} R.,  et~al., 2018, \mn@doi [\apj] {10.3847/1538-4357/aac6cb}, \href
  {http://adsabs.harvard.edu/abs/2018ApJ...860..124D} {860, 124}

\bibitem[\protect\citeauthoryear{{Foreman-Mackey}, {Hogg}, {Lang}  \&
  {Goodman}}{{Foreman-Mackey} et~al.}{2013}]{emcee2013PASP..125..306F}
{Foreman-Mackey} D.,  {Hogg} D.~W.,  {Lang} D.,   {Goodman} J.,  2013, \mn@doi
  [\pasp] {10.1086/670067}, \href
  {http://adsabs.harvard.edu/abs/2013PASP..125..306F} {125, 306}

\bibitem[\protect\citeauthoryear{{Gaia Collaboration} et~al.,}{{Gaia
  Collaboration} et~al.}{2016}]{Gaia2016A&A...595A...1G}
{Gaia Collaboration} et~al., 2016, \mn@doi [\aap]
  {10.1051/0004-6361/201629272}, \href
  {http://adsabs.harvard.edu/abs/2016A%26A...595A...1G} {595, A1}

\bibitem[\protect\citeauthoryear{{Goodman} \& {Weare}}{{Goodman} \&
  {Weare}}{2010}]{MCMC2010CAMCS...5...65G}
{Goodman} J.,  {Weare} J.,  2010, \mn@doi [Communications in Applied
  Mathematics and Computational Science, Vol.~5, No.~1, p.~65-80, 2010]
  {10.2140/camcos.2010.5.65}, \href
  {http://adsabs.harvard.edu/abs/2010CAMCS...5...65G} {5, 65}

\bibitem[\protect\citeauthoryear{{Grady} et~al.,}{{Grady}
  et~al.}{2013}]{Grady2013ApJ...762...48G}
{Grady} C.~A.,  et~al., 2013, \mn@doi [\apj] {10.1088/0004-637X/762/1/48},
  \href {http://adsabs.harvard.edu/abs/2013ApJ...762...48G} {762, 48}

\bibitem[\protect\citeauthoryear{{Isella}, {P{\'e}rez}, {Carpenter}, {Ricci},
  {Andrews}  \& {Rosenfeld}}{{Isella} et~al.}{2013}]{Isella2013ApJ...775...30I}
{Isella} A.,  {P{\'e}rez} L.~M.,  {Carpenter} J.~M.,  {Ricci} L.,  {Andrews}
  S.,   {Rosenfeld} K.,  2013, \mn@doi [\apj] {10.1088/0004-637X/775/1/30},
  \href {http://adsabs.harvard.edu/abs/2013ApJ...775...30I} {775, 30}

\bibitem[\protect\citeauthoryear{{Isella}, {Chandler}, {Carpenter}, {P{\'e}rez}
   \& {Ricci}}{{Isella} et~al.}{2014}]{Isella2014ApJ...788..129I}
{Isella} A.,  {Chandler} C.~J.,  {Carpenter} J.~M.,  {P{\'e}rez} L.~M.,
  {Ricci} L.,  2014, \mn@doi [\apj] {10.1088/0004-637X/788/2/129}, \href
  {http://adsabs.harvard.edu/abs/2014ApJ...788..129I} {788, 129}

\bibitem[\protect\citeauthoryear{{Kataoka}, {Okuzumi}, {Tanaka}  \&
  {Nomura}}{{Kataoka} et~al.}{2014}]{Kataoka2014A&A...568A..42K}
{Kataoka} A.,  {Okuzumi} S.,  {Tanaka} H.,   {Nomura} H.,  2014, \mn@doi [\aap]
  {10.1051/0004-6361/201323199}, \href
  {http://adsabs.harvard.edu/abs/2014A%26A...568A..42K} {568, A42}

\bibitem[\protect\citeauthoryear{{Koller}, {Li}  \& {Lin}}{{Koller}
  et~al.}{2003}]{Koller2003ApJ...596L..91K}
{Koller} J.,  {Li} H.,   {Lin} D.~N.~C.,  2003, \mn@doi [\apjl]
  {10.1086/379032}, \href {http://adsabs.harvard.edu/abs/2003ApJ...596L..91K}
  {596, L91}

\bibitem[\protect\citeauthoryear{{Lyra} \& {Klahr}}{{Lyra} \&
  {Klahr}}{2011}]{LyraKlahr2011A&A...527A.138L}
{Lyra} W.,  {Klahr} H.,  2011, \mn@doi [\aap] {10.1051/0004-6361/201015568},
  \href {http://adsabs.harvard.edu/abs/2011A%26A...527A.138L} {527, A138}

\bibitem[\protect\citeauthoryear{{Lyra} \& {Lin}}{{Lyra} \&
  {Lin}}{2013}]{LyraLin2013ApJ...775...17L}
{Lyra} W.,  {Lin} M.-K.,  2013, \mn@doi [\apj] {10.1088/0004-637X/775/1/17},
  \href {http://adsabs.harvard.edu/abs/2013ApJ...775...17L} {775, 17}

\bibitem[\protect\citeauthoryear{{Lyra}, {Johansen}, {Zsom}, {Klahr}  \&
  {Piskunov}}{{Lyra} et~al.}{2009}]{Lyra2009A&A...497..869L}
{Lyra} W.,  {Johansen} A.,  {Zsom} A.,  {Klahr} H.,   {Piskunov} N.,  2009,
  \mn@doi [\aap] {10.1051/0004-6361/200811265}, \href
  {http://adsabs.harvard.edu/abs/2009A%26A...497..869L} {497, 869}

\bibitem[\protect\citeauthoryear{{Mac{\'{\i}}as}, {Anglada}, {Osorio},
  {Torrelles}, {Carrasco-Gonz{\'a}lez}, {G{\'o}mez}, {Rodr{\'{\i}}guez}  \&
  {Sierra}}{{Mac{\'{\i}}as} et~al.}{2017}]{Macias2017ApJ...838...97M}
{Mac{\'{\i}}as} E.,  {Anglada} G.,  {Osorio} M.,  {Torrelles} J.~M.,
  {Carrasco-Gonz{\'a}lez} C.,  {G{\'o}mez} J.~F.,  {Rodr{\'{\i}}guez} L.~F.,
  {Sierra} A.,  2017, \mn@doi [\apj] {10.3847/1538-4357/aa6620}, \href
  {http://adsabs.harvard.edu/abs/2017ApJ...838...97M} {838, 97}

\bibitem[\protect\citeauthoryear{{Marino}, {Casassus}, {Perez}, {Lyra},
  {Roman}, {Avenhaus}, {Wright}  \& {Maddison}}{{Marino}
  et~al.}{2015}]{Marino2015ApJ...813...76M}
{Marino} S.,  {Casassus} S.,  {Perez} S.,  {Lyra} W.,  {Roman} P.~E.,
  {Avenhaus} H.,  {Wright} C.~M.,   {Maddison} S.~T.,  2015, \mn@doi [\apj]
  {10.1088/0004-637X/813/1/76}, \href
  {http://adsabs.harvard.edu/abs/2015ApJ...813...76M} {813, 76}

\bibitem[\protect\citeauthoryear{{Mittal} \& {Chiang}}{{Mittal} \&
  {Chiang}}{2015}]{Mittal2015ApJ...798L..25M}
{Mittal} T.,  {Chiang} E.,  2015, \mn@doi [\apjl]
  {10.1088/2041-8205/798/1/L25}, \href
  {http://adsabs.harvard.edu/abs/2015ApJ...798L..25M} {798, L25}

\bibitem[\protect\citeauthoryear{{Miyake} \& {Nakagawa}}{{Miyake} \&
  {Nakagawa}}{1993}]{MiyakeNakagawa1993Icar..106...20M}
{Miyake} K.,  {Nakagawa} Y.,  1993, \mn@doi [\icarus] {10.1006/icar.1993.1156},
  \href {http://adsabs.harvard.edu/abs/1993Icar..106...20M} {106, 20}

\bibitem[\protect\citeauthoryear{{Muto} et~al.,}{{Muto}
  et~al.}{2015}]{Muto2015PASJ...67..122M}
{Muto} T.,  et~al., 2015, \mn@doi [\pasj] {10.1093/pasj/psv098}, \href
  {http://adsabs.harvard.edu/abs/2015PASJ...67..122M} {67, 122}

\bibitem[\protect\citeauthoryear{{P{\'e}rez}, {Isella}, {Carpenter}  \&
  {Chandler}}{{P{\'e}rez} et~al.}{2014}]{Perez_L_2014ApJ...783L..13P}
{P{\'e}rez} L.~M.,  {Isella} A.,  {Carpenter} J.~M.,   {Chandler} C.~J.,  2014,
  \mn@doi [\apjl] {10.1088/2041-8205/783/1/L13}, \href
  {http://adsabs.harvard.edu/abs/2014ApJ...783L..13P} {783, L13}

\bibitem[\protect\citeauthoryear{{Ragusa}, {Dipierro}, {Lodato}, {Laibe}  \&
  {Price}}{{Ragusa} et~al.}{2017}]{Ragusa2017MNRAS.464.1449R}
{Ragusa} E.,  {Dipierro} G.,  {Lodato} G.,  {Laibe} G.,   {Price} D.~J.,  2017,
  \mn@doi [\mnras] {10.1093/mnras/stw2456}, \href
  {http://adsabs.harvard.edu/abs/2017MNRAS.464.1449R} {464, 1449}

\bibitem[\protect\citeauthoryear{{Rau} \& {Cornwell}}{{Rau} \&
  {Cornwell}}{2011}]{RauCornwell2011A&A...532A..71R}
{Rau} U.,  {Cornwell} T.~J.,  2011, \mn@doi [\aap]
  {10.1051/0004-6361/201117104}, \href
  {http://adsabs.harvard.edu/abs/2011A%26A...532A..71R} {532, A71}

\bibitem[\protect\citeauthoryear{{Reg{\'a}ly}, {Juh{\'a}sz}, {S{\'a}ndor}  \&
  {Dullemond}}{{Reg{\'a}ly} et~al.}{2012}]{Regaly2012MNRAS.419.1701R}
{Reg{\'a}ly} Z.,  {Juh{\'a}sz} A.,  {S{\'a}ndor} Z.,   {Dullemond} C.~P.,
  2012, \mn@doi [\mnras] {10.1111/j.1365-2966.2011.19834.x}, \href
  {http://adsabs.harvard.edu/abs/2012MNRAS.419.1701R} {419, 1701}

\bibitem[\protect\citeauthoryear{{Reggiani} et~al.,}{{Reggiani}
  et~al.}{2014}]{Reggiani2014ApJ...792L..23R}
{Reggiani} M.,  et~al., 2014, \mn@doi [\apjl] {10.1088/2041-8205/792/1/L23},
  \href {http://adsabs.harvard.edu/abs/2014ApJ...792L..23R} {792, L23}

\bibitem[\protect\citeauthoryear{{S{\'a}ndor}, {Lyra}  \&
  {Dullemond}}{{S{\'a}ndor} et~al.}{2011}]{Sandor2011ApJ...728L...9S}
{S{\'a}ndor} Z.,  {Lyra} W.,   {Dullemond} C.~P.,  2011, \mn@doi [\apjl]
  {10.1088/2041-8205/728/1/L9}, \href
  {http://adsabs.harvard.edu/abs/2011ApJ...728L...9S} {728, L9}

\bibitem[\protect\citeauthoryear{{Shakura} \& {Sunyaev}}{{Shakura} \&
  {Sunyaev}}{1973}]{1973A&A....24..337S}
{Shakura} N.~I.,  {Sunyaev} R.~A.,  1973, \aap, \href
  {http://adsabs.harvard.edu/abs/1973A%26A....24..337S} {24, 337}

\bibitem[\protect\citeauthoryear{{Sierra}, {Lizano}  \& {Barge}}{{Sierra}
  et~al.}{2017}]{Sierra2017ApJ...850..115S}
{Sierra} A.,  {Lizano} S.,   {Barge} P.,  2017, \mn@doi [\apj]
  {10.3847/1538-4357/aa94c1}, \href
  {http://adsabs.harvard.edu/abs/2017ApJ...850..115S} {850, 115}

\bibitem[\protect\citeauthoryear{{Varni{\`e}re} \& {Tagger}}{{Varni{\`e}re} \&
  {Tagger}}{2006}]{Varniere2006A&A...446L..13V}
{Varni{\`e}re} P.,  {Tagger} M.,  2006, \mn@doi [\aap]
  {10.1051/0004-6361:200500226}, \href
  {http://adsabs.harvard.edu/abs/2006A%26A...446L..13V} {446, L13}

\bibitem[\protect\citeauthoryear{{Weidenschilling}}{{Weidenschilling}}{1977}]{Weidenschilling1977MNRAS.180...57W}
{Weidenschilling} S.~J.,  1977, \mnras, \href
  {http://adsabs.harvard.edu/abs/1977MNRAS.180...57W} {180, 57}

\bibitem[\protect\citeauthoryear{{Wright} \& {Barlow}}{{Wright} \&
  {Barlow}}{1975}]{WrightBarlow1975MNRAS.170...41W}
{Wright} A.~E.,  {Barlow} M.~J.,  1975, \mn@doi [\mnras]
  {10.1093/mnras/170.1.41}, \href
  {http://adsabs.harvard.edu/abs/1975MNRAS.170...41W} {170, 41}

\bibitem[\protect\citeauthoryear{{Zhu} \& {Baruteau}}{{Zhu} \&
  {Baruteau}}{2016}]{ZB2016MNRAS.458.3918Z}
{Zhu} Z.,  {Baruteau} C.,  2016, \mn@doi [\mnras] {10.1093/mnras/stw202}, \href
  {http://adsabs.harvard.edu/abs/2016MNRAS.458.3918Z} {458, 3918}

\bibitem[\protect\citeauthoryear{{Zhu} \& {Stone}}{{Zhu} \&
  {Stone}}{2014}]{Zhu_Stone_2014ApJ...795...53Z}
{Zhu} Z.,  {Stone} J.~M.,  2014, \mn@doi [\apj] {10.1088/0004-637X/795/1/53},
  \href {http://adsabs.harvard.edu/abs/2014ApJ...795...53Z} {795, 53}

\bibitem[\protect\citeauthoryear{{de Val-Borro}, {Artymowicz}, {D'Angelo}  \&
  {Peplinski}}{{de Val-Borro} et~al.}{2007}]{deVal-Borro2007A&A...471.1043D}
{de Val-Borro} M.,  {Artymowicz} P.,  {D'Angelo} G.,   {Peplinski} A.,  2007,
  \mn@doi [\aap] {10.1051/0004-6361:20077169}, \href
  {http://adsabs.harvard.edu/abs/2007A%26A...471.1043D} {471, 1043}

\bibitem[\protect\citeauthoryear{{van der Marel} et~al.,}{{van der Marel}
  et~al.}{2013}]{vanderMarel2013Sci...340.1199V}
{van der Marel} N.,  et~al., 2013, Science, \href
  {http://adsabs.harvard.edu/abs/2013Sci...340.1199V} {340, 1199}

\bibitem[\protect\citeauthoryear{{van der Marel}, {van Dishoeck}, {Bruderer},
  {P{\'e}rez}  \& {Isella}}{{van der Marel}
  et~al.}{2015a}]{vandermarel_2015A&A...579A.106V}
{van der Marel} N.,  {van Dishoeck} E.~F.,  {Bruderer} S.,  {P{\'e}rez} L.,
  {Isella} A.,  2015a, \mn@doi [\aap] {10.1051/0004-6361/201525658}, \href
  {http://adsabs.harvard.edu/abs/2015A%26A...579A.106V} {579, A106}

\bibitem[\protect\citeauthoryear{{van der Marel}, {Pinilla}, {Tobin}, {van
  Kempen}, {Andrews}, {Ricci}  \& {Birnstiel}}{{van der Marel}
  et~al.}{2015b}]{vandermarel_2015ApJ...810L...7V}
{van der Marel} N.,  {Pinilla} P.,  {Tobin} J.,  {van Kempen} T.,  {Andrews}
  S.,  {Ricci} L.,   {Birnstiel} T.,  2015b, \mn@doi [\apjl]
  {10.1088/2041-8205/810/1/L7}, \href
  {http://adsabs.harvard.edu/abs/2015ApJ...810L...7V} {810, L7}

\bibitem[\protect\citeauthoryear{{van der Marel}, {van Dishoeck}, {Bruderer},
  {Andrews}, {Pontoppidan}, {Herczeg}, {van Kempen}  \& {Miotello}}{{van der
  Marel} et~al.}{2016a}]{vdMarel_2016A&A...585A..58V}
{van der Marel} N.,  {van Dishoeck} E.~F.,  {Bruderer} S.,  {Andrews} S.~M.,
  {Pontoppidan} K.~M.,  {Herczeg} G.~J.,  {van Kempen} T.,   {Miotello} A.,
  2016a, \mn@doi [\aap] {10.1051/0004-6361/201526988}, \href
  {http://adsabs.harvard.edu/abs/2016A%26A...585A..58V} {585, A58}

\bibitem[\protect\citeauthoryear{{van der Marel}, {Cazzoletti}, {Pinilla}  \&
  {Garufi}}{{van der Marel} et~al.}{2016b}]{vdM2016ApJ...832..178V}
{van der Marel} N.,  {Cazzoletti} P.,  {Pinilla} P.,   {Garufi} A.,  2016b,
  \mn@doi [\apj] {10.3847/0004-637X/832/2/178}, \href
  {http://adsabs.harvard.edu/abs/2016ApJ...832..178V} {832, 178}

\bibitem[\protect\citeauthoryear{{van der Plas}, {M{\'e}nard}, {Canovas},
  {Avenhaus}, {Casassus}, {Pinte}, {Caceres}  \& {Cieza}}{{van der Plas}
  et~al.}{2017}]{vdP2017A&A...607A..55V}
{van der Plas} G.,  {M{\'e}nard} F.,  {Canovas} H.,  {Avenhaus} H.,  {Casassus}
  S.,  {Pinte} C.,  {Caceres} C.,   {Cieza} L.,  2017, \mn@doi [\aap]
  {10.1051/0004-6361/201731392}, \href
  {http://adsabs.harvard.edu/abs/2017A%26A...607A..55V} {607, A55}

\makeatother
\end{thebibliography}

\bsp	% typesetting comment
\label{lastpage}
\end{document}